\documentclass[11pt]{amsart}

\usepackage{amssymb}
\usepackage{amsmath,amsthm,amsfonts}  
 \usepackage{graphicx}
\usepackage{amssymb}
\usepackage{amsmath}
\usepackage{amsthm}
\usepackage{pdfsync}
\usepackage{fancyhdr}
\usepackage[applemac]{inputenc}

\usepackage[default]{frcursive}
\usepackage[T1]{fontenc}

\usepackage{color}
\usepackage{xcolor}
\usepackage[breaklinks,colorlinks,backref]{hyperref}
\hypersetup{
    colorlinks, 
    linktoc=all, 
    linkcolor=red,  
  citecolor=hyptxt,
  urlcolor=blue}
  \hypersetup{
  citebordercolor=,
  filebordercolor=green,
  linkbordercolor=blue
}
\definecolor{hyptxt}{rgb}{0.7, 0.4, 0.9}
\usepackage{bibtopic}

\usepackage[full]{textcomp} 
     \usepackage{fbb} 
     \usepackage[scaled=.95]{cabin}
     \usepackage[varqu,varl]{inconsolata}
     \usepackage[libertine,bigdelims,vvarbb]{newtxmath}
     \usepackage[cal=boondoxo]{mathalfa}
     \usepackage[T1]{fontenc}
\useosf

\definecolor{hervecolor}{rgb}{0.8,0,0.7}
     
\newcommand{\ket}[1]{|\kern.3ex#1\kern.3ex\rangle}
\newcommand{\bra}[1]{\langle\kern.3ex #1 \kern.3ex|}
\newcommand{\scalar}[2]{\langle\kern.3ex #1 \kern.3ex|\kern.3ex#2\kern.3ex\rangle}

\newcommand{\ii}{\mathsf{i}}
\newcommand{\btr}{\blacktriangleright}

\def\R{\mathbb{R}}
\def\N{\mathbb{N}}
\def\C{\mathbb{C}}

\def\lg{\langle }
\def\rg{\rangle }
\def\deq{\stackrel{\mathrm{def}}{=}}
\def\adg{a^{\dag}}
\def\vp{\varphi}
\def\vap{\varpi}

\def\vs{\varsigma}

\def\ud{\mathrm{d}}

\def\sfM{\mathsf{M}}
\def\sfP{\mathsf{P}}

\def\mfF{\mathfrak{F}}

\def\mFs{\mathfrak{F_s}}
\def\mcs{\mathfrak{c_s}}
\def\omFs{\overline{\mFs}}
\def\mfs{\mathfrak{f_s}}
\def\omfs{\overline{\mathfrak{f_s}}}
\def\Da{\mbox{\LARGE \textit{a}}}
\def\Dd{\mbox{\Large\textit{d}}}
\def\Dc{\mbox{\LARGE \textit{c}}}
\def\TDc{\widetilde{\mbox{\LARGE \textit{c}}}}
\def\TDa{\widetilde{\mbox{\LARGE \textit{a}}}}
\def\TDd{\widetilde{\mbox{\Large\textit{d}}}}

\numberwithin{equation}{section}

\begin{document}
\date{\today}
 
\title[WH quantization compendium]{Weyl-Heisenberg integral  quantization(s): \\a compendium }
\author[Bergeron, Curado, Gazeau, Rodrigues]{
H. Bergeron$^{\mathrm{a}}$, 
E.M.F.  Curado$^{\mathrm{b,c}}$,  \\
J.-P. Gazeau$^{\mathrm{b,d}}$, 
 Ligia M.C.S. Rodrigues$^{\mathrm{b}}$}
\address{\emph{$^{\mathrm{a}}$
Univ Paris-Sud, ISMO, UMR 8214, 91405 Orsay, France}}
\address{\emph{$^{\mathrm{b}}$ Centro Brasileiro de Pesquisas Fisicas } \\
\emph{   $^{\mathrm{c}}$ Instituto Nacional de Ci\^encia e Tecnologia - Sistemas Complexos}\\
\emph{  Rua Xavier Sigaud 150, 22290-180 - Rio de Janeiro, RJ, Brazil  }} 
\address{\emph{  $^{\mathrm{d}}$ APC, UMR 7164,}\\
\emph{ Univ Paris  Diderot, Sorbonne Paris Cit\'e,  }  
\emph{75205 Paris, France}} 

\email{e-mail: herve.bergeron@u-psud.fr, 
 evaldo@cbpf.br,
gazeau@apc.univ-paris7.fr, ligia@cbpf.br}

{\abstract{We present a list of formulae  useful for Weyl-Heisenberg integral quantizations, with arbitrary weight, of functions or distributions on the plane. Most of these formulae are known, others are original. The list encompasses particular cases like Weyl-Wigner quantization (constant weight) and  coherent states (CS) or Berezin quantization (Gaussian weight). The formulae are given  with implicit assumptions on their validity on appropriate space(s) of functions (or distributions). One of the aims of the document is to accompany a work in progress on Weyl-Heisenberg integral quantization of dynamics for the motion of a point particle on the line.  
 }}

\maketitle

\tableofcontents

\section{Introduction}
\label{intro}
While preparing  a paper on the Weyl-Heisenberg (WH) integral quantization  of one dimensional  dynamics for the motion of a point particle on the real line \cite{bercugaro16}, we have been establishing a compendium of useful formulae. We now think  suitable to share these results with the community of researchers interested in similar topics. Most of the presented material is not original. It can be found in many places, and with different notations, since the emergence of Quantum Mechanics \cite{vanderWaer68}, specially in References \cite{born_jordan25} (see also \cite{fedak09}), \cite{born_heis_jord26}, \cite{dirac25}, \cite{weyl28}, \cite{dirac30} (see also  \cite{dirac82}), \cite{mccoy32}, \cite{groenewold46}, \cite{moyal47}, \cite{vanhove51}, \cite{dirac64}, \cite{cohen66}, \cite{cohenbook12}, \cite{AgaWo70},  in the mathematically oriented  \cite{howe80} and \cite{folland87}, in \cite{ozorio88}, in the books \cite{zachos06} and \cite{gazeaubook09}, in \cite{bergayou13}, \cite{bergaz14}...    For most recent, mathematically oriented, works, see for instance \cite{aagbook14}, \cite{aniello15} and \cite{cordero_etal15}, and references therein. However, we do not pretend to have exhausted here the complete relevant bibliography!

Despite that lack of originality of many formulas presented in our paper, the fact to gather all of them in a single document, under an itemized form,  might be valuable for people working on the subject. Beyond their computational justifications, one might be concerned by the mathematical validity of our expressions in terms of functional analysis, since most of them should be justified on a mathematical level with regard to involved (generalized) functions. Nevertheless, they are given  with implicit assumptions on their validity on appropriate space(s) of functions (or distributions). 

The object we start with, namely the Euclidean plane $\R^2$, or the complex plane $\C$, is briefly presented in Section 
\ref{phasespace} together with its physical content. Weyl-Heisenberg  group and algebra together with their representations  as operators in Hilbert spaces are described in Section \ref{WHgroup},   Section \ref{WHalgebrarep}, and  Section \ref{WHUIR} respectively. Section \ref{D(z)} is devoted to the so-called WH displacement operator, $D(z)$ (complex notation), $\mathcal{D}(q,p)$ (real notations), of central importance for the content of this paper. The short section \ref{CSstandard} is devoted to the construction of standard coherent states through the action of the $D(z)$ on the ``vacuum'' state in Hilbert space.  Section \ref{PTsym} deals with discrete symmetries and  in Section \ref{rotations} are considered rotations in the plane. Integral formulas involving the displacement operator $D$ or $\mathcal{D}$ are given in Section \ref{intformD(z)}.  Section \ref{Fourier} is devoted to symplectic Fourier analysis of (possibly operator valued) functions on the plane.  Trace formulae are presented in Section \ref{trace}. The heart of the paper lies in  Section \ref{WHquantization} where is presented what we call Weyl-Heisenberg integral quantization (WHIQ), expressed in either  complex or real variables, and the ingredients are just the WH displacement operator and a weight ($\sim$  apodization) function. In the succeeding sections are given technical issues of this quantization procedure, in relation with derivatives in Section \ref{quantderiv}, with products in Section \ref{quantprod}, expansion coefficients in    Sections \ref{expcoefI}, \ref{expcoefII},  Section \ref{expcoefIII}, and  Section \ref{expcoefIV}.  Common characteristics and outcomes of WHIQ are presented in Section \ref{permissues}.  Section \ref{partweights} is devoted to WHIQ results obtained from particular weight or measures on the phase space.  We examine in Section \ref{sepquant} the quantization of separable functions $f(q,p) = u(q)v(p)$. In  Section \ref{comrules} we give a list of commutation relations which are relevant to examples encountered in previous Sections, and in Section \ref{gausssepquant} the case of weights which are separable gaussians with different widths.  Section \ref{QOintq} is devoted to the WHIQ versions of the harmonic oscillator and their common outcomes. Finally, semiclassical portraits, generalizing Wigner or Husimi functions for operators issued from WHIQ are considered in  Section \ref{wignerandothers}, with an interesting outcome in terms of probabilistic interpretation of the weight function. 

\section{Phase space}
\label{phasespace}
\begin{itemize}
  \item Phase space for motion on the line is $\R^2 \sim \C$
\begin{equation}
\label{phaspace1}
\C \sim \R^2= \left\{z = \frac{q+\ii p}{\sqrt{2}}\, , \,q\,, \,  p \in \R\right\}\, . 
\end{equation}
\begin{equation}
\label{measurephs}
\ud^2 z = \frac{\ud q \,\ud p}{2}
\end{equation}
 \item Physical dimensions are restored, particularly  in view of classical limit,  through the re-scalings
\begin{equation}
\label{physdim}
\C \ni z = \frac{q+\ii p}{\sqrt{2}} \mapsto z= \frac{1}{\sqrt{2}}\left(\frac{q}{\ell} + \ii \,\frac{p}{\wp}\right)\, , 
\end{equation}
so that the complex $z$ remains dimensionless.
\item Here $\ell$ (resp. $\wp$) is some length (resp. momentum) appropriate to the scale of the model. Thinking to quantum  systems, we can also introduce  the Planck constant $\hbar$ such that $\ell\wp = \hbar$    
\item Useful change of variables formulae 
\begin{itemize}
  \item With physical constants $\hbar$ and $\ell$
  \begin{align}
\label{pdzbzqp}
 \partial_{z}   &= \frac{1}{\sqrt{2}} \left(\ell \partial_q - \ii \frac{\hbar}{\ell}\partial_p\right)\, , \quad \partial_{\bar z}   = \frac{1}{\sqrt{2}} \left(\ell \partial_q + \ii \frac{\hbar}{\ell}\partial_p\right)\, ,   \\
 \label{pdqpzbz}   \partial_{q}   &= \frac{1}{\sqrt{2}\,\ell} \left(\partial_{z} +  \partial_{\bar z} \right)  \, , \quad \partial_{p}   = \frac{\ii \ell}{\sqrt{2}\,\hbar} \left(\partial_{z} -  \partial_{\bar z} \right)\, .  
    \end{align}
  \item Without physical constants
    \begin{align}
\label{zbzqp}
 \partial_{z}   &= \frac{1}{\sqrt{2}} \left(\partial_q - \ii \partial_p\right)\, , \quad \partial_{\bar z}   = \frac{1}{\sqrt{2}} \left( \partial_q + \ii \partial_p\right)\, ,  \\
 \label{qpzbz}   \partial_{q}   &= \frac{1}{\sqrt{2}} \left(\partial_{z} +  \partial_{\bar z} \right)  \, , \quad \partial_{p}   = \frac{\ii}{\sqrt{2}} \left(\partial_{z} -  \partial_{\bar z} \right) \, . 
    \end{align}
\end{itemize}
\end{itemize}

\section{Weyl-Heisenberg  group}
\label{WHgroup}
\begin{itemize}
  \item Forgetting about physical dimensions and $\hbar = 1$, an arbitrary  
element $g$ of  $\mathrm{G}_{\rm WH}$ is of the form   
\begin{equation}
\label{weylheispar}
 G_{\mathrm{WH}}= \{ g = (\varsigma , q,p)\, , \, \vs \in {\mathbb R}\,,\,  (q,p)  \in  {\mathbb R}^{2}\} \, .
\end{equation} 
 \item In complex notations
  \begin{equation}
\label{weylheisG}
\nonumber G_{\mathrm{WH}}= \{(\vs,z)\, , \, \vs\in \R , \, z\in \C\}\, . 
\end{equation} 
\item Multiplication law  
\begin{equation}
\begin{aligned}
g_{1}g_{2} &= (\vs_{1} + \vs_{2} + \xi ((q_{1},p_{1}) \, ,\, (q_{2},p_{2}))\, ,   
           q_{1}+q_{2},\; p_{1}+p_{2}) \\
           & = (\vs_{1} + \vs_{2} + \xi(z_1,z_2)\, ,  
           z_1 + z_2)\, ,  
\label{WHgroupmult} 
\end{aligned}
\end{equation}
\item where $\xi$ is the multiplier  (or two-cocycle) function ($\sim$ symplectic form on $\R^2$)
\begin{equation}
\label{ximultip}
\xi ((q_{1},p_{1})\, ,\, (q_{2},p_{2})) = \frac 1{2} (p_{1}q_{2} - p_{2}q_{1})\equiv \xi(z_1,z_2)=  \mathrm{Im}\,z_1\bar z_2 \equiv -z_1\wedge z_2 \, . 
\end{equation}
\item Besides  this  multiplicity of notations, we introduce another the further one, more suitable for many expressions
\begin{equation}
\label{defcirc}
z_1\circ z_2 := z_1\bar z_2 - \bar z_1 z_2 = 2\xi(z_1,z_2)= 2 \mathrm{Im}\,z_1\bar z_2\, .
\end{equation}
\item The  two-cocycle condition results from associativity of the WH group
\begin{equation}
\label{cocycle}
\xi(z_1,z_2) + \xi(z_1+z_2,z_3) = \xi(z_2,z_3) +\xi(z_1,z_2 + z_3)
\end{equation}
together with $\xi(z,0)= 0= \xi(0,z)$ and $\xi(z,-z)= 0$, resulting  from group identity and inverse respectively. 
\end{itemize}

\section{Weyl-Heisenberg algebra and its Fock or number representation}
\label{WHalgebrarep}
\begin{itemize}
\item Notational convention: set of nonnegative integers is $\N= 0,1, 2 , \dotsc$.
\item Let $\mathcal{H}$ be a separable (complex) Hilbert space  with orthonormal basis $e_0,e_1,\dots, e_n \equiv |e_n \rangle, \dots$, (e.g. the Fock space with $|e_n\rg \equiv |n\rg$). 

\item Define the lowering and raising operators $a$ and $a^\dag$ as
\begin{align}
   a\, \ket{e_n} & = \sqrt{n} \ket{e_{n-1}}\, , \quad  a\ket{e_0} = 0 \, \quad \mbox{(lowering or annihilation operator)}\\
   a^{\dag} \, \ket{e_n} & = \sqrt{n +1} \ket{e_{n+1}} \quad \mbox{(raising or creation operator)}\, .
\end{align}  
\item Equivalently
\begin{align}
   a\,  & =\sum_{n=0}^\infty \sqrt{n+1} \ket{e_n} \bra{e_{n+1}}  \quad \mbox{(weak sense)} \, \\
   a^{\dag} \,  & = \sum_{n=0}^\infty \sqrt{n+1} \ket{e_{n+1}} \bra{e_{n}}  \quad \mbox{(weak sense)}\, . 
\end{align}  
  \item Operator algebra $\{a,\adg, I\}$ is defined by the Canonical Commutation Rule (CCR)
  \begin{equation}
\label{ccr}
  [a,\adg]= I\,.
\end{equation}
  \item Number operator: $N= \adg a$, spectrum $\N$, $N|e_n\rg = n |e_n\rg$.  
  \item Operators  $Q$ and $P$
  \begin{equation}
\label{opQopP}
a= \frac{Q+\ii P}{\sqrt{2}}\, , \quad \adg= \frac{Q-\ii P}{\sqrt{2}} \Leftrightarrow Q= \frac{a + \adg}{\sqrt{2}}\, , \quad P= \frac{1}{\ii}\frac{a-\adg}{\sqrt{2}}\, . 
\end{equation}
\item Both are essentially self-adjoint in $\mathcal{H}$, with absolutely spectrum $\sigma(Q)= \R= \sigma(P)$ 
 \item Familiar form of the CCR
 \begin{equation}
\label{CCRQP}
[Q,P]= \ii\,I\, . 
\end{equation}
\item Consistently to \eqref{CCRQP},  if $Q$ is realised on a multiplication operator on the Hilbert space $L^2(\R,\ud x)$ of square-integrable complex valued functions (``wave functions'')  on its spectrum $\sigma(Q)= \R$, $Q\phi(x)=x\phi(x)$, then $P$ is realised as $-\ii \dfrac{\ud\hphantom{x}}{\ud x}$
\item Alternatively, if $P$ is realised on a multiplication operator on the Hilbert space $L^2(\R,\ud k)$ of square-integrable complex valued functions  on its spectrum $\sigma(P)= \R$, $P\hat \phi(k)=k\hat\phi(k)$ then $Q$ is realised as $\ii \dfrac{\ud\hphantom{x}}{\ud k}$
\end{itemize}

\section{Unitary Weyl-Heisenberg group representation(s) }
\label{WHUIR}
\begin{itemize}
\item Any infinite-dimensional UIR, $U^{\lambda}$, of $\mathrm{G}_{\rm WH}$ is characterized by a real number $\lambda \neq 0$
(in addition, there are also degenerate, one-dimensional, UIRs corresponding to $\lambda = 0$) and can be identified with an irreducible  
representation of the CCR  
\begin{equation}
(\vs,q,p)\mapsto U^{\lambda}(\vs, q, p)     
                             :=  e^{\ii \lambda(\vs- pq/2)}
                             e^{\ii \lambda pQ} e^{- \ii \lambda qP}. 
\label{WHgrouprep1} 
\end{equation}
\item If the  Hilbert space carrying the UIR is $\mathcal{H} = L^{2}({\mathbb R} , \ud x)$, $\R$ being viewed here as the spectrum $\sigma(Q)$  of the (essentially) self-adjoint position operator $Q$,  the unitary operator $U^{\lambda}$ acts as 
\begin{equation}
(U^{\lambda}(\vs , q, p)\phi )(x) = e^{\ii \lambda \vs}
    e^{\ii \lambda p(x- q/2)}\phi (x-q)\,, \qquad  
    \phi \in L^{2}({\mathbb R} \, , {\ud}x)\, . 
\label{WHgrouprep2} 
\end{equation}
\item Alternatively,  if the  Hilbert space carrying the UIR is $\mathcal{H} = L^{2}({\mathbb R} , \ud x)$, $\R$ being viewed here as the spectrum$\sigma(P)$  of the (essentially) self-adjoint position operator $P$,  the unitary operator $U^{\lambda}$ acts as 
\begin{equation}
\label{WHgrouprep3} 
(U^{\lambda}(\vs , q, p)\hat\phi )(k) = e^{\ii \lambda \vs}
    e^{-\ii \lambda q(k- p/2)}\hat\phi (k-p)\,, \qquad  
    \hat\phi \in L^{2}({\mathbb R} \, , {\ud}k)\, . 
\end{equation}
\item Thus, the  infinitesimal generators  of  $U^{\lambda}$ read as
\begin{equation} 
\begin{split}
\mathsf D^{\lambda}_\vs&=\ii \lambda I\, , \quad  (\mathsf D^{\lambda}_p\phi)(x) =  \ii \lambda x\phi (x)\equiv \ii \lambda Q\phi(x) \, ,\\   (\mathsf D^{\lambda}_q\phi)(x) & = 
   -\frac{\partial\phi}{\partial x}(x) =-\ii\lambda P\phi(x)\,,\quad
   [\mathsf D^{\lambda}_p,\mathsf D^{\lambda}_q] = \ii \lambda=  \lambda^2[Q,P]\,.
\label{genrepWHgroup}
\end{split}
\end{equation}
\item Usually, one takes for $\lambda$ the specific value, 
$\lambda =1/\hbar$ (in QM) or simply  $\lambda=1$, and  write $U$ for $U^1$
\item Hence, the representation reads as the action of exponential operators
\begin{equation}
\label{repWHQP}
(U(\vs , q, p )= e^{\ii \vs}
    e^{-\ii pq/2} \, e^{\ii pQ}\, e^{-\ii qP} \equiv e^{\ii \vs}\, \mathcal{D}(q,p)
\end{equation}
where $\mathcal{D}(q,p)$ is the unitary ``displacement operator'' in representation $(q,p)$
\item Now consider  the Fock-Bargmann Hilbert space  $\mathcal{FB}$ (resp. $\mathcal{AFB}$) of entire analytical (resp. antianalytical) functions   that are square integrable with respect to the Gaussian measure on the complex plane
  \begin{equation}
\label{gaussmeas}
 \frac{1}{\pi }e^{-\vert z \vert^2}\, \ud^2z = \frac{1}{2\pi}e^{-\frac{q^2 +p^2}{2}}\, \ud q\,\ud p\, ,
\end{equation} 
  \begin{equation}
\label{convFB}
\nonumber \alpha(z) \ (\mathrm{resp.} \ \alpha(\bar z)) = \sum_{n = 0}^{+\infty} \alpha_n z^n\ (\mathrm{resp.} \ \bar z^n) \ \mbox{converges absolutely for all}\ z \in \C \,   ,
\end{equation}
i.e., its convergence radius is infinite, and 
  \begin{equation}
\label{convFB1}
 \Vert \alpha \Vert^2_{\mathcal{FB}} \deq    \int_{\C} \vert \alpha(z) \vert^2 \, e^{-\vert z \vert^2}\, \frac{\ud^2z}{\pi} < \infty\, .
\end{equation}
\item Space  $\mathcal{FB}$  is equipped with the scalar product:
\begin{equation}
\label{FBscpr}
\lg \alpha_1|\alpha_2\rg =  \int_{\C} \overline{\alpha_1(z)}  \alpha_2(z)  \,  = \sum_{n = 0}^{+\infty} n! \, \overline{\alpha_{1n}}\, \alpha_{2n}\, . 
\end{equation}
\item The unitary operator $U$ acts on space  $\mathcal{AFB}$ as
\begin{equation}
\label{UAFB}
(U(\vs , z^{\prime}) \alpha )(\bar z)= e^{\ii  \vs}\,e^{-\frac{1}{2}\vert z^{\prime}\vert^2}\, e^{z^{\prime}\bar z}\,\alpha(\bar z-\bar z^{\prime})= 
e^{\ii \lambda \vs}\,e^{-\frac{1}{2}\vert z- z^{\prime}\vert^2}\, e^{\frac{\vert z\vert^2}{2}}\,e^{\frac{z^{\prime}\circ z}{2}}\,\alpha(\bar z-\bar z^{\prime})\, , 
\end{equation}
where $z^{\prime}\circ z= -z\circ z^{\prime}$ is defined in \eqref{defcirc}
\item Equivalently, and more simply, on the  space of ``Gaussian weighted $\mathcal{AFB}$ as
\begin{equation}
\label{UGWAFB}
(U(\vs , z^{\prime}) \alpha_g )(\bar z)= 
e^{\ii \lambda \vs}\,e^{\frac{z^{\prime}\circ z}{2}}\,\alpha_g(\bar z-\bar z^{\prime})\, , \quad \alpha_g (z) = e^{-\frac{\vert z\vert^2}{2}}\,\alpha(z) \in L^2\left(\C,\frac{\ud^2 z}{\pi}\right)\,.
\end{equation}
\item On Space $\mathcal{FB}$ the annihilation operator $a$ is represented as a derivation whereas its adjoint is a multiplication operator:
\begin{equation}
\label{FBaadag}
 a\, \alpha(z) =  \frac{\ud}{\ud z} \alpha(z)\, , \qquad 
  \adg\, \alpha(z) =z\, \alpha(z)\, .
\end{equation}

\end{itemize} 

\section{Displacement operator}
\label{D(z)}
\begin{itemize}
\item From the previous section, to each complex number $z$ is associated the (unitary) displacement operator  or ``function $D(z)$'' 
\begin{equation}
\label{displac}
\C \ni z \mapsto D(z) = e^{z\adg -\bar z a}\, ,\quad D(-z) = (D(z))^{-1} = D(z)^{\dag}\, . 
\end{equation}
\item In variables $(q,p)$  and operators $Q$ and $P$
\begin{equation}
\label{dispopQP}
D(z)\equiv \mathcal{D}(q,p)= e^{\ii (pQ-qP)}\,. 
\end{equation}
 \item Unitary representation $U$ with complex notations 
  \begin{align}
\label{unrepWH}
(\vs,z) &\mapsto e^{\ii\vs}D(z)\, , \\ (\vs,z)(\vs^{\prime},z^{\prime}) &\mapsto e^{\ii\vs}D(z)e^{\ii\vs^{\prime}}D(z^{\prime}) = e^{\ii(\vs+\vs^{\prime} + \mathrm{Im}\, z\bar z^{\prime})} D(z+z^{\prime})\,. 
\end{align}
\item The  Weyl  formula 
\begin{equation}
\label{bacahau}
e^A \,e^B = e^{ \frac{1}{2} [A,B]} \, e^{(A + B)}\, ,
\end{equation}
(arising from the Baker-Campbell-Hausdorff relation) which is formally\footnote{Actually we have to be seriously aware of domains of involved operators when we apply the ``algebraic'' formula, see \cite{fraga16} } valid for any pair of operators that  commute with their commutator, $ [A,[A,B]] = 0 =  [B,[A,B]] $, yields 
\begin{equation}
\label{camhaus}
D(z)=e^{-\frac{1}{2} |z|^2}\, e^{z a^\dag} e^{-\bar{z} a} =e^{\frac{1}{2} |z|^2}\, e^{-\bar{z} a} e^{z a^\dag}\, ,
\end{equation}
\item In variables $(q,p)$  and operators $Q$ and $P$, consistently with \eqref{repWHQP}
\begin{equation}
\label{camhausQP}
D(z) \equiv \mathcal{D}(q,p)= e^{\ii (pQ-qP)}= e^{-\ii\frac{ qp}{2}}\,e^{\ii pQ}\,e^{-\ii qP}= e^{\ii\frac{ qp}{2}}\,e^{-\ii qP}\,e^{\ii pQ}  \,. 
\end{equation}
\item It follows the formulae
\begin{equation}
\label{propdisp1}
\dfrac{\partial}{\partial z} D(z) = \left(a^\dag - \dfrac{1}{2} \bar{z} \right) D(z) = D(z) \left( a^\dag + \dfrac{1}{2} \bar{z} \right).
\end{equation}
\begin{equation}
\label{propdisp2}
\dfrac{\partial}{\partial \bar{z}} D(z) = - \left(a - \dfrac{1}{2} z \right) D(z) = - D(z) \left( a + \dfrac{1}{2} z\right).
\end{equation}
\item equivalently
\begin{equation}
\label{propdisp3}
z\, D(z) = [a,D(z)]\,, \qquad \bar z\, D(z) = [a^{\dag},D(z)]\, , 
\end{equation}
\begin{equation}
\label{propdisp4}
\dfrac{\partial}{\partial z} D(z) = \frac{1}{2}\{\adg,D(z)\}\, , \qquad \dfrac{\partial}{\partial \bar{z}} D(z) =  -\frac{1}{2}\{a,D(z)\}\, .
\end{equation}
\item With variables $q$, $p$
\begin{equation}
\label{propdispqp1}
\dfrac{\partial}{\partial q} \mathcal{D}(q,p) = \left(-\ii \,P + \dfrac{\ii}{2} \,p \right) \mathcal{D}(q,p) = -\mathcal{D}(q,p) \left( \ii P + \dfrac{\ii}{2}\,p  \right)\, , 
\end{equation}
\begin{equation}
\label{propdispqp2}
\dfrac{\partial}{\partial p} \mathcal{D}(q,p) =  \left(\ii \, Q - \dfrac{\ii}{2} \,q \right) \mathcal{D}(q,p) =  \mathcal{D}(q,p) \left( \ii\,Q +  \dfrac{\ii}{2}\,q \right)\, ,
\end{equation}
\item equivalently
\begin{equation}
\label{propdispqp3}
q\, \mathcal{D}(q,p) = [Q,\mathcal{D}(q,p)]\,, \qquad p\, \mathcal{D}(q,p) = [P,\mathcal{D}(q,p)]\, , 
\end{equation}
\begin{equation}
\label{propdispqp4}
\dfrac{\partial}{\partial q} \mathcal{D}(q,p) = -\ii\,\{P,\mathcal{D}(q,p)\}\, , \qquad \dfrac{\partial}{\partial p} \mathcal{D}(q,p) =  \ii\,\{Q,\mathcal{D}(q,p)\}\, .
\end{equation}

\item Addition formula
\begin{align}
\label{additiveDz}
D(z)D(z^{\prime}) & = e^{\frac{1}{2}z \circ z^{\prime}}D(z+z^{\prime}) \,,\\ 
\label{additiveDqp} \mathcal{D}(q,p)\,\mathcal{D}(q^{\prime},p^{\prime})&= e^{-\frac{\ii}{2}(qp^{\prime}-pq^{\prime})}\mathcal{D}(q+q^{\prime},p+p^{\prime})\, .
\end{align}
\item It follows the covariance formula on a global level
\begin{align}
\label{Dcovglobz}
D(z) D(z^{\prime}) D(z)^\dag &= e^{z \circ z^{\prime}} D(z^{\prime})\, , \\
\label{Dcovglobqp}\mathcal{D}(q,p)\,\mathcal{D}(q^{\prime},p^{\prime})\,\mathcal{D}^{\dag}(q,p)&=e^{\ii ( pq^{\prime} -qp^{\prime})} \mathcal{D}(q^{\prime},p^{\prime})\,. 
\end{align}
\item and on a Lie algebra level
\begin{equation}
\label{Dcovinf}
 \quad D(z) a D(z)^\dag= a - zI, \quad D(z) a^\dag D(z)^\dag= a^\dag - \bar{z}I\, .  
\end{equation}
\item Matrix elements of operator $D(z)$ in the basis $\{|e_n\rg\}$ involve associated Laguerre polynomials $L^{(\alpha)}_n(t)$ \cite{magnus66,gradryz07}:
\begin{equation}
\label{matelD}
\lg e_m|D(z)|e_n\rg = D_{m n}(z) =  \sqrt{\dfrac{n!}{m!}}\,e^{-\vert z\vert^{2}/2}\,z^{m-n} \, L_n^{(m-n)}(\vert z\vert^{2})\, ,   \quad \mbox{for} \ m\geq n\, , 
\end{equation}
with  $L_n^{(m-n)}(t) = \frac{m!}{n!} (-t)^{n-m}L_m^{(n-m)}(t)$ for $n\geq m$ 
\item Orthonormality properties straightforwardly derive from unitarity:
\begin{equation}
\label{unitortho}
\int_{\C}\frac{\ud^2 z}{\pi}\, D_{mn}(z)\, \overline{D_{m'n'}(z)}= \delta_{mm'}\delta_{nn'}\,.
\end{equation}
\item One derives from unitarity the infinite sums:
\begin{equation}
\label{sumser}
\sum_{n=0}^{\infty}D_{mn}(z)\overline{D_{m'n}(z)}= \delta_{mm'}= \sum_{n=0}^{\infty}D_{nm'}(z)\overline{D_{nm}(z)}\, ,
\end{equation}
and particularly
\begin{equation}
\label{sumser2}
\sum_{n=0}^{\infty} \vert D_{mn}(z)\vert^2= 1= \sum_{n=0}^{\infty} \vert D_{nm}(z)\vert^2 \, , \quad m \in \N\, .
\end{equation}
\end{itemize}

\section{Coherent states}
\label{CSstandard}
\begin{itemize}
  \item Standard (Schr\"{o}dinger \cite{schrodinger26}, Iwata \cite{iwata51}, Klauder \cite{klauder60,klauder63a,klauder63b}, Glauber \cite{glauber63a,glauber63b,glauber63c}, Sudarshan \cite{sudarshan63}, see also \cite{klauskagbook85,perelomov86,zhangfenggil90,dodonov02,dodoman03,gazeaubook09,aagbook14} for reviews and further developments),  coherent states are defined as  vectors in $\mathcal{H}$
  \begin{equation}
\label{scs}
|z\rg = D(z)\,|e_0\rg = e^{-\frac{\vert z\vert^2}{2}}\sum_{n=0}^{\infty} \frac{z^n}{\sqrt{n!}}\,|e_n\rg\,, 
\end{equation}
  \item They have unit norm and solve the identity in $\mathcal{H}$
  \begin{equation}
\label{resunit}
\lg z |z\rg = 1\, , \qquad  \int_{\C}\frac{\ud^2 z}{\pi} |z\rg\lg z| = I\, . 
\end{equation}
  \item Their overlap is a Gaussian on $\C$ up to the symplectic phase factor
  \begin{equation}
\label{overlapcs}
\lg z|z^{\prime}\rg = e^{\frac{1}{2}z^{\prime}\circ z}\,e^{\frac{1}{2}\vert z-z^{\prime}\vert^2}\, .
\end{equation}
  \item CS is   reproducing kernel for  functions $\psi(z):=\lg z|\psi\rg$ built from vectors $|\psi \rg \in \mathcal{H}$
  \begin{equation}
\label{reprker}
\psi(z)=  \int_{\C}\frac{\ud^2 z^{\prime}}{\pi}  \lg z | z^{\prime}\rg \, \psi(z^{\prime})\, .  
\end{equation}
\item In representation position or momentum,  there are Gaussian states, expressed with $z=\frac{q+\ii p}{\sqrt{2}}$, 
\begin{equation}
\label{CSx}
\lg x|e_0\rg= \frac{1}{\sqrt[4]{\pi}}e^{-\frac{x^2}{2}}\, , \quad \lg x|z\rg = \frac{1}{\sqrt[4]{\pi}}\,e^{\ii p(x-q/2)}\,e^{-\frac{(x-q)^2}{2}}\, , 
\end{equation}
\begin{equation}
\label{CSk}
\lg k |e_0\rg= \frac{1}{\sqrt[4]{\pi}}e^{-\frac{k^2}{2}}\, , \quad \lg k|z\rg = \frac{1}{\sqrt[4]{\pi}}\,e^{-\ii q(k-p/2)}\,e^{-\frac{(k-p)^2}{2}}\, . 
\end{equation}
\item WH translation covariance of coherent states result from their definition  \eqref{scs} as ``displaced lowest state'' 
\begin{equation}
\label{covcs}
D(z)\,|z_0\rg = e^{\frac{1}{2}z\circ z_0}\, |z+z_0\rg\,. 
\end{equation}

\end{itemize}

\section{Parity and time reversal}
\label{PTsym}
\begin{itemize}
  \item The parity ${\sf P}$ acts on $\mathcal{H}$ as a linear operator through
\begin{equation}
\label{parity}
{\sf P} \ket{e_n}= (-1)^n \ket{e_n}\, , \quad \mbox{or}\quad {\sf P}= e^{\ii \pi a^\dag a}\,. 
\end{equation}
 \item The time reversal ${\sf T}$ acts on $\mathcal{H}$ as a conjugation, that is an {\it antilinear operator} such that
\begin{equation}
\label{timerev}
{\sf T} \sum_{n}\xi_n\ket{e_n}= \sum_n \overline{\xi_n}\ket{e_n}\, .
\end{equation}
 \item These discrete symmetries verify
\begin{align}
 {\sf P}^2 &={\sf T}^2 =I, \\
 {\sf P} a {\sf P}& =-a \, , \qquad {\sf P} a^\dag {\sf P} = -a^\dag,\\
  {\sf T} a {\sf T}& =a \, , \qquad{\sf T} a^\dag {\sf T} = a^\dag,\\
  {\sf P} D(z) {\sf P}&=D(- z) \, , \qquad {\sf T} D(z) {\sf T} = D(\bar{z}).
 \end{align}
\end{itemize}

\section{Rotation in the plane}
\label{rotations}
\begin{itemize}
  \item  A unitary representation $\theta \mapsto U_{\mathbb{T}}(\theta)$ of the torus $\mathbb{S}^1$ on the Hilbert space $\mathcal{H}$ is defined as 
\begin{equation}
\label{unrotplane}
U_{\mathbb{T}}(\theta)|e_n\rg = e^{\ii (n + \nu) \theta}|e_n\rg\, , \quad \nu \in \R\, .
\end{equation}
\item Note that ${\sf P}= U_{\mathbb{T}}(\pi)$ with $\nu = 0$
 \item Rotational covariance of the displacement operator
\begin{equation}
\label{rotcovD}
U_{\mathbb{T}}(\theta)D(z)U_{\mathbb{T}}(\theta)^{\dag} = D\left(e^{\ii \theta}z\right)\, . 
\end{equation}
 \item Rotational covariance of coherent states
\begin{equation}
\label{rotcovcs}
U_{\mathbb{T}}(\theta)\,|z\rg = e^{\ii \nu\theta}\,\left|e^{\ii \theta}z\right\rg\, . 
\end{equation}

\end{itemize}
\section{Integral formulae for $D(z)$}
\label{intformD(z)}
\begin{itemize}

\item First  fundamental integral: from \cite{magnus66}
\begin{equation}
\label{fundintD}
\int_0^{\infty} e^{-\frac{t}{2}}\, L_n(t)\, \ud t  = (-2)^n\, \Rightarrow \int_{\C}  D_{m n}(z)\, \frac{\ud^2 z}{\pi}= \delta_{mn} (-2)^m\,,
\end{equation}
it follows 
\begin{equation}
\label{fourD0}
\int_{\C} D(z) \, \frac{\ud^2 z}{\pi}= 2{\sf P}\, .
\end{equation}
\item Resolution of the identity  follows from (\ref{fourD0})
\begin{equation}
\label{resunweyl}
\int_{\C}  D(z)\, 2{\sf P}\, D(-z)\,\frac{\ud^2 z}{\pi}  = I\, . 
\end{equation}
(At the basis of the Weyl-Wigner quantization (in complex notations), see below)
\item Second fundamental integral: from (\ref{matelD}) and  the orthogonality of the associated Laguerre polynomials we obtain the ``ground state'' projector as the Gaussian average of $D(z)$
\begin{equation}
\label{lapD}
 \int_{\mathbb{C}} e^{- \frac{1}{2} |z|^2} D(z)\, \dfrac{\ud^2 z}{\pi}= \ket{e_0} \bra{e_0} \,. 
\end{equation}
\item More generally for $\mathrm{Re}(s) < 1$
\begin{equation}
\label{lapDs}
\int_{\mathbb{C}} e^{\frac{s}{2} |z|^2} D(z)\,  \dfrac{\ud^2 z}{\pi} = \dfrac{2}{1-s} \exp \left( \ln \dfrac{s+1}{s-1} a^\dag a \right)\,, 
\end{equation}
where the convergence holds in norm for $\mathrm{Re}(s)<0$ and weakly for $0 \leq \mathrm{Re}(s) < 1$.
 \end{itemize}

\section{Harmonic analysis on $\C$ or $\R^2$ and symbol calculus}
\label{Fourier}
\subsection{In terms of $z$ and $\bar z$}
\begin{itemize}
  \item Symplectic Fourier transform on $\mathbb{C}$ (for a sake of simplicity, we write $f(z,\bar z) \equiv f(z)$)
  \begin{align}
 \label{symFourTr1}  \mathfrak{f_s}[f](z)&=\int_{\mathbb{C}} e^{ z \bar \xi -\bar z \xi} f(\xi)\,  \frac{\ud^2 \xi}{\pi}= \int_{\mathbb{C}} e^{2\ii \,\mathrm{Im}(z \bar \xi)} f(\xi)\,  \frac{\ud^2 \xi}{\pi}\\
\label{symFourTr2}   &=\int_{\mathbb{C}} e^{z\circ\xi} f(\xi)\,  \frac{\ud^2 \xi}{\pi}\quad \mbox{(notation most used in the paper)}\, . 
     \end{align}
\item  Dirac-Fourier formula
\begin{equation}
\label{diracfourier}
\mathfrak{f_s}[1](z)= \int_{\mathbb{C}}e^{z \circ \xi }\, \dfrac{\ud^2 \xi}{\pi} = \int_{\R^2}e^{-\ii (qy-px)}\, \dfrac{\ud x\,\ud y}{2\pi} =  2\pi \delta(q)\,\delta(p)=\pi \delta^{2}(z)\, .
\end{equation}
\item The symplectic Fourier transform is its inverse: it is an involution
\begin{equation}
\label{ff1}
\mathfrak{f_s}[\mathfrak{f_s}[f]](z) = f(z) \  \Leftrightarrow \ \mathfrak{f_s}\,\mathfrak{f_s} = \mathfrak{f_s}^2 = I\, .
\end{equation}
\item The symplectic Fourier transform commutes with the parity operator
\begin{equation}
\label{invsymFourTr2}
\mfs= \sfP \,\mfs \,\sfP \, , \quad (\sfP\,f)(z)=f(-z)= \tilde{f}(z)\,, \ \tilde{f}(z):= f(-z)\, . 
\end{equation}
\item Reflected symplectic Fourier transform
\begin{equation}
\label{refsymFourTr1} \overline{\mathfrak{f_s}}[f](z)=\int_{\mathbb{C}} e^{-z\circ\xi} f(\xi)\,  \frac{\ud^2 \xi}{\pi}=  \mathfrak{f_s}[f](-z)=  \mathfrak{f_s}\left[\tilde{f}\right](z)= \overline{\mathfrak{f_s}\left[\bar{f}\right](z)}\, . 
\end{equation}
\item The reflected symplectic Fourier transform is its inverse
\begin{equation}
\label{rfrf1}
\omfs \,\omfs = I \,. 
\end{equation}
\item Factorization of the parity operator
\begin{equation}
\label{factP}
\omfs\mfs = \mfs \omfs = \sfP\,. 
\end{equation}

\item Symplectic Fourier transform and translation\\
with 
\begin{equation}
\label{translation}
(\mathsf{t}_z\,f)(z^{\prime}):= f(z^{\prime}-z)\, , 
\end{equation}
\begin{equation}
\label{symfourtransl}
\mfs\left[\mathsf{t}_z\,f\right](z^{\prime})= e^{z^{\prime}\circ z}\, \mfs[f](z^{\prime})\,,\quad \omfs\left[\mathsf{t}_{-z}\,f\right](z^{\prime})= e^{z^{\prime}\circ z}\, \omfs[f](z^{\prime}) \,. 
\end{equation}
\item Symplectic Fourier transform and derivation
\begin{align}
\label{symFourder1}
   \frac{\partial^k}{\partial z^k} \,\mfs[f](z)&= \mfs\left[\bar\xi^k\, f\right](z)\, ,  \quad  &\frac{\partial^k}{\partial z^k} \,\omfs[f](z)= \omfs\left[\left(-\bar\xi\right)^k\, f\right](z) \, ,     \\
 \label{symFourder2}    \frac{\partial^k}{\partial \bar z^k} \,\mfs[f](z)&= \mfs\left[(-\xi)^k\,f\right](z)\, , \quad &\frac{\partial^k}{\partial \bar z^k} \,\omfs[f](z)= \omfs\left[\xi^k\,f\right](z)\, ,   \\
 \label{symFourder3}  \mfs\left[\frac{\partial^k}{\partial \xi^k} \,f\right](z)&= \bar z^k\,\mfs\left[f\right](z)\, , \quad &\omfs\left[\frac{\partial^k}{\partial \xi^k} \,f\right](z)=(- \bar z)^k\,\omfs\left[f\right](z)\, ,  \\
 \label{symFourder4}   \mfs\left[\frac{\partial^k}{\partial \bar\xi^k} \,f\right](z)&= (-z)^k\,\mfs\left[f\right](z)\, , \quad &\omfs\left[\frac{\partial^k}{\partial \bar\xi^k} \,f\right](z)=z^k\,\omfs\left[f\right](z)\, . 
\end{align}
\item Convolution product with complex variables
\begin{equation}
\label{convolC}
(f\ast g)(z):= \int_{\C}\ud^2z^{\prime}\, f(z-z^{\prime}) \, g(z^{\prime})= (g\ast f)(z)\, . 
\end{equation}
\item Symplectic Fourier transform of convolution products
\begin{align}
\label{symfourconv1}
  \mfs[f\ast g] (z)&= \pi \,  \mfs[f] (z)\, \mfs[ g] (z)\, ,  \\
 \label{symfourconv2}    \mfs[f\, g] (z)&= \frac{1}{\pi} \,  (\mfs[f] \ast \mfs[ g]) (z)\, . 
\end{align}
\item Symplectic Fourier transform of Gaussian
\begin{equation}
\label{symfourgauss}
\mfs\left[e^{\nu \, \vert\xi\vert^2}\right](z) = \frac{1}{(-\nu)}\, e^{ \frac{\vert z\vert^2}{\nu}}= \omfs\left[e^{\nu \, \vert\xi\vert^2}\right](z)\, , \quad \mathrm{Re}(\nu) < 0\, . 
\end{equation}
 \item Symplectic Fourier transform of  $D$  
  \begin{equation}
\label{ftransfD}
\int_{\C} e^{z \circ z^{\prime}} \, D(z^{\prime})\,\frac{\ud^2 z^{\prime}}{\pi}  = 2\, D(2z)\,{\sf P} = 2 {\sf P}\, D(-2z)  \,.
\end{equation}
\end{itemize}
\subsection{In terms of $q$ and $p$}
\begin{itemize}
  \item In terms of coordinates $z= (q+\ii p)/\sqrt{2}$, $\xi= (x+\ii y)/\sqrt{2}$,
\begin{equation}
\label{symFourqp}
\mathfrak{f_s}[f](z)\equiv \mathfrak{F_s}[F](q,p)= \int_{\R^2}e^{-\ii (qy - px)}\, F(x,y)\,\frac{\ud x\,\ud y}{2\pi} = \mathfrak{F}[F](-p,q)\, , 
\end{equation}
where $ \mathfrak{F}$ denotes the standard two-dimensional Fourier transform,
\begin{equation}
\label{stFourqp}
\mathfrak{F}[F](k_x,k_y)= \int_{\R^2}e^{-\ii (k_x x + k_y y)}\, F(x,y)\,\frac{\ud x\,\ud y}{2\pi}\, , 
\end{equation}
with inverse
\begin{equation}
\label{stFourqpinv}
\overline{\mathfrak{F}}[F](k_x,k_y)= \int_{\R^2}e^{\ii (k_x x + k_y y)}\, F(x,y)\,\frac{\ud x\,\ud y}{2\pi}= \mathfrak{F}[F](-k_x,-k_y)\, .
\end{equation} 
  \item $\mathfrak{F_s}$ is involutive, $\mathfrak{F_s}\left[\mathfrak{F_s}[F]\right]= \mathfrak{F_s}^2[F]= F$ like its ``dual'' defined as
  \begin{equation}
\label{dsymFourqp}
\overline{\mathfrak{F_s}}[F](q,p)= \mathfrak{F_s}[F](-q,-p)=\int_{\R^2}e^{\ii (qy - px)}\, F(x,y)\,\frac{\ud x\,\ud y}{2\pi} = \mathfrak{F}[F](p,-q)\, , 
\end{equation}
 \item Symplectic Fourier transform and derivation
\begin{align}
\label{symFourderqp1}
   \frac{\partial^k}{\partial q^k} \,\mFs[F](q,p)&= (-\ii)^k \mFs\left[y^k\, F\right](q,p)\, ,    &\frac{\partial^k}{\partial q^k} \,\omFs[F](q,p)= \ii^k\mFs\left[y^k\, F\right](q,p) \, ,     \\
 \label{symFourderqp2}   \frac{\partial^k}{\partial p^k} \,\mFs[F](q,p)&= \ii^k\mFs\left[x^k\, F\right](q,p)\, ,  &\frac{\partial^k}{\partial p^k} \,\omFs[F](q,p)= (-\ii)^k\omFs\left[x^k\,F\right](q,p)\, ,   \\
 \label{symFourderqp3}  \ii^k\mFs\left[\frac{\partial^k}{\partial x^k} \,F\right](q,p)&= p^k\,\mfs\left[F\right](q,p)\, ,  &(-\ii)^k\omFs\left[\frac{\partial^k}{\partial x^k} \,f\right](q,p)=p^k\,\omFs\left[F\right](q,p)\, ,  \\
 \label{symFourderqp4}  (-\ii)^k \mFs\left[\frac{\partial^k}{\partial y^k} \,F\right](q,p)&= q^k\,\mFs\left[F\right](q,p)\, , &\ii^k\omFs\left[\frac{\partial^k}{\partial y^k} \,F\right](q,p)=q^k\,\omFs\left[F\right](q,p)\, . 
\end{align}
\item Convolution product 
\begin{equation}
\label{convolC}
(F\ast G)(q,p):= \int_{\C}\ud q^{\prime}\,\ud p^{\prime}\, F(q-q^{\prime}, p-p^{\prime}) \, G(q-q^{\prime},p-p^{\prime})= (G\ast F)(z)\, . 
\end{equation}
\item Symplectic Fourier transform of convolution products
\begin{align}
\label{symfourconvqp1}
  \mFs[F\ast G] (q,p)&= 2\pi \,  \mFs[F] (q,p)\, \mFs[ G] (q,p)\, ,  \\
 \label{symfourconvqp2}    \mFs[F\, G] (q,p)&= \frac{1}{2\pi} \,  (\mFs[F] \ast \mFs[ G]) (q,p)\, . 
\end{align}
\item Same formulae for $\omFs$
\end{itemize}
\subsection{In terms of $r$ and $\phi$, polar coordinates of $z$}
\begin{itemize}
\item  Consider a function (or distribution) $f(z)$ with $z=r\,e^{\ii \phi}$ Fourier expandable as
\begin{equation}
\label{fourserfz}
f\left(r\,e^{\ii \phi}\right) =  \sum_{n=-\infty}^{+\infty}c_n(r)\,e^{\ii n\phi}\, .
\end{equation}
  \item Its symplectic Fourier transform is expressed as the Fourier series
  \begin{equation}
\label{fsrphi}
\mfs[f]\left(r\,e^{\ii \phi}\right) = \sum_{n=-\infty}^{+\infty}\mcs_n(r)\,e^{\ii n\phi}\, ,
\end{equation}
where the coefficients are  Hankel transform of order $n$ \cite{magnus66}, up to factors $2$,  of the $c_n$'s
\begin{equation}
\label{hankel}
\mcs_n(r) = 2\int_0^{\infty}c_n(\rho)\,J_n(2r\rho)\,\rho\,\ud \rho\, , 
\end{equation}
($J_n$: Bessel function of the first kind, with $J_{-n} = (-1)^n\,J_n$)
\end{itemize}
 \section{Trace formulae for $D(z)$}
 \label{trace}
\begin{itemize}
\item From the resolution of the identity of the standard coherent states,  from Eq. (\ref{Dcovglobz}) and Eq. (\ref{diracfourier}),  it follows
\begin{equation}
\label{traceD}
\mathrm{tr} \, D(z) = \int_{\C}  \langle \xi | D(z) | \xi \rangle\,  \frac{\ud^2 \xi}{\pi}  = \pi \delta^2(z) \,.
\end{equation}
\item Using Eq.(\ref{additiveDz}), we have
\begin{equation}
\label{traceDD1}
\mathrm{tr} \left(D(z)^\dag D(z^{\prime}) \right) = \pi \delta^2(z-z^{\prime}) \,.
\end{equation}
\end{itemize}

\section{Noncommutative Weyl-Heisenberg harmonic analysis}
\begin{itemize}
  \item Weyl-Heisenberg transform of a function as the operator in $\mathcal{H}$
  \begin{equation}
\label{WHtransform}
f(z) \mapsto \mathfrak{D}[f] = \int_{\C}D(z)\, f(z)\,\frac{\ud^2 z}{\pi}\, . 
\end{equation}
  \item Adjoint
  \begin{equation}
\label{adjDf}
\mathfrak{D}[f]^{\dag} = \mathfrak{D}[\tilde f]\,.  
\end{equation}
  \item Inversion formula is direct consequence of \eqref{traceDD1}
  \begin{equation}
\label{ncWHinversion}
f(z) = \mathrm{tr}  \, (D(-z)\, \mathfrak{D}[f])\, . 
\end{equation}
\item Translation covariance
\begin{equation}
\label{trcovWHtr}
D(z_0)\mathfrak{D}[f]D(z_0)^{\dag}= \mathfrak{D}\left[\mathsf{t}_zf\right]
\end{equation}
\end{itemize}

\section{Weyl-Heisenberg integral quantization}
\label{WHquantization}
\begin{itemize}
\item Pick a \textit{weight} function $\varpi(z) \equiv \Pi(q,p)$  obeying 
\begin{equation}
\label{varpi0}
\varpi(0) = 1= \Pi(0,0)\, . 
\end{equation}
\item Suppose that 
\begin{equation}
\label{opMvarpi}
{\sf M}^{\vap}= \int_{\mathbb{C}} D(z)\,  \varpi(z)\, \frac{\ud^2 z}{\pi}=\int_{\R^2} \mathcal{D}(q,p)\,  \Pi(q,p)\, \frac{\ud q \ud p}{2\pi}\equiv M^{\Pi}\, .
\end{equation}
is bounded on $\mathcal{H}$
\item Then, the family 
\begin{equation}
\label{dispMz}
{\sf M}^{\vap}(z):= D(z) {\sf M}^{\vap}D(z)^\dag  =\mathcal{D}(q,p){\sf M}^{\Pi} \mathcal{D}(q,p)^{\dag}= {\sf M}^{\Pi}(q,p)
\end{equation}
resolves the identity on $\mathcal{H}$
\begin{equation}
\label{residMz}
\int_{\mathbb{C}} \, {\sf M}^{\vap}(z) \,\frac{\ud^2 z}{\pi}= I = \int_{\R^2} {\sf M}^{\Pi} (q,p)\,\frac{\ud q \ud p}{2\pi}\, . 
\end{equation}
\item The operator ${\sf M}^{\vap}$ as an integral operator on Gaussian weighted anti-analytic Fock-Bargmann space
\begin{equation}
\label{MintopAFB}
\left({\sf M}^{\vap}\alpha_g\right)(\bar z) = \int_{\C}\frac{\ud^2 z^{\prime}}{\pi}\, \mathfrak{M}^{\vap}(z,z^{\prime})\,\alpha_g(\bar z^{\prime})\, , \quad \alpha_g \in L^2\left(\C,\frac{\ud^2 z}{\pi}\right)\, ,
\end{equation}
where $\alpha_g(\bar z):= e^{-\frac{\vert z\vert^2}{2}}\,\alpha(\bar z)$, $\alpha\in \mathcal{AFB}$. The  kernel $\mathfrak{M}^{\vap}$ is given by
\begin{equation}
\label{MkerAFB}
\mathfrak{M}^{\vap}(z,z^{\prime})= \vap(z-z^{\prime})\,e^{\frac{z\circ z^{\prime}}{2}}\,.
\end{equation}
\item In terms of  symplectic Fourier transforms,
\begin{equation}
\label{MAFBSF}
\left({\sf M}^{\vap}\alpha_g\right)(\bar z) = \frac{1}{\pi}e^{\cdot \circ z}\omfs[\vap]\ast\mfs[\alpha_g]\left(\frac{\bar z}{2}\right)\, . 
\end{equation}
\item The operator ${\sf M}^{\vap}$ as an integral operator in representation position
\begin{equation}
\label{Mintop}
\left({\sf M}^{\vap}\phi\right)(x) = \int_{-\infty}^{+\infty}\ud x^{\prime}\, \mathcal{M}^{\vap}(x,x^{\prime})\,\phi(x^{\prime})\, , \quad \phi \in L^2(\R,\ud x)\, ,
\end{equation}
with kernel
\begin{equation}
\label{Mker}
\mathcal{M}^{\vap}(x,x^{\prime})= \frac{1}{\sqrt{2\pi}} \, \hat\Pi_p\left(x-x^{\prime},-\frac{x+x^{\prime}}{2}\right)\,.
\end{equation}
\item In the above expression, $\hat\Pi_p$ stands for the partial Fourier transform of $\Pi(q,p)$ with respect to the $p$ variable
\begin{equation}
\label{partFour}
\hat\Pi_p(q,y)= \frac{1}{\sqrt{2\pi}}\int_{-\infty}^{+\infty}\ud p \, e^{-\ii yp}\, \Pi(q,p)\, .
\end{equation}
\item A trace formula, issued, if applicable, from \eqref{traceD} and  \eqref{varpi0}
\begin{equation}
\label{traceMom}
\mathrm{tr}\left({\sf M}^{\vap}(z)\right)=\mathrm{tr}\left({\sf M}^{\vap}\right)= \vap(0)=1\, . 
\end{equation}
\item Necessary condition on $\vap(z)$ for that ${\sf M}^{\vap}(z)$ define a normalized Positive Operator Valued Measure (POVM)
\begin{equation}
\label{MomPOVM}
\forall\, z\, ,  \  0< \lg z| {\sf M}^{\vap} |z\rg  =  \omfs\left[ e^{-\frac{\vert \xi\vert^2}{2}} \, \vap(\xi)\right](z) = \frac{2}{\pi}\,\omfs\left[ e^{-\frac{\vert \xi\vert^2}{2}} \, \right]\ast \omfs\left[\vap(\xi)\right](z)\, .
\end{equation}
\item Weyl-Heisenberg integral quantization is the linear map
\begin{equation}
\label{eqquantvarpi}
f \mapsto A^{\vap}_f = \int_{\mathbb{C}} \, {\sf M}^{\vap}(z) \, \, f(z) \,\frac{\ud^2 z}{\pi}\, , 
\end{equation}
such that the constant function $f=1$ is mapped to the  identity $I$. 
\item Alternatively with  the symplectic Fourier transform \eqref{symFourTr2}
\begin{equation}
\label{quantvarpi1}
A^{\vap}_f = \int_{\mathbb{C}}  D(z)\, \mathfrak{f_s}[f](-z)\, \varpi(z) \,\frac{\ud^2 z}{\pi} = \int_{\mathbb{C}}  D(z)\, \omfs[f](z)\, \varpi(z) \,\frac{\ud^2 z}{\pi}\, .
\end{equation}
\item In terms of variables $q$, $p$ and Fourier transform, with $\varpi(z) \equiv \Pi(q,p)$,
\begin{align}
\label{quantPi1}
A^{\vap}_f \equiv A^{\Pi}_F&= \int_{\R^2}  \mathcal{D}(q,p)\, \mFs[F](-q,-p)\, \Pi(q,p) \,\frac{\ud q\, \ud p}{2\pi}\\ 
 \label{quantPi1b}&= \int_{\R^2}  \mathcal{D}(q,p)\, \overline{\mFs}[F](q,p)\, \Pi(q,p) \,\frac{\ud q\, \ud p}{2\pi}\\ 
 \label{quantPi2}&= \int_{\R^2} \int_{\R^2} e^{-\frac{\ii qp}{2}}\,e^{\ii pQ}\,e^{-\ii qP} \, e^{\ii (qy - px)}\, F(x,y)\, \Pi(q,p) \,\frac{\ud q\, \ud p}{2\pi}  \, \frac{\ud x\, \ud y}{2\pi} \\
\label{quantPi3}&= \int_{\R^2} \int_{\R^2} e^{\frac{\ii qp}{2}}\,e^{-\ii qP}\,e^{\ii pQ} \, e^{\ii (qy - px)}\, F(x,y)\, \Pi(q,p) \,\frac{\ud q\, \ud p}{2\pi}  \, \frac{\ud x\, \ud y}{2\pi}
\end{align}
\item The operator $A^{\vap}_f$ as an integral operator on Gaussian weighted anti-analytic Fock-Bargmann space
\begin{equation}
\label{Mzaintop}
\left(A^{\vap}_f\alpha_g\right)(\bar z) = \int_{\C}\frac{\ud^2 z^{\prime}}{\pi}\, \mathfrak{A}_f^{\vap}(z,z^{\prime})\,\alpha_g(\bar z^{\prime})\, ,
\end{equation}
with kernel
\begin{equation}
\label{Afzker}
 \mathfrak{A}_f^{\vap}(z,z^{\prime})=  \vap(z-z^{\prime})\,e^{\frac{z\circ z^{\prime}}{2}}\,\omfs[f](z-z^{\prime})= \mathfrak{M}^{\vap}(z,z^{\prime})\,\omfs[f](z-z^{\prime})\,.
\end{equation}
\item In terms of  symplectic Fourier transforms,
\begin{equation}
\label{AfAFBSF}
\left(A^{\vap}_f\alpha_g\right)(\bar z) = \frac{1}{\pi^2}e^{\cdot \circ z}\,f\ast e^{\cdot \circ z}\omfs[\vap]\ast\mfs[\alpha_g]\left(\frac{\bar z}{2}\right)\, . 
\end{equation}
\item The operator $A^{\vap}_f=A^{\Pi}_F$ as an integral operator in representation position
\begin{equation}
\label{Mzfintop}
\left(A^{\Pi}_F\phi\right)(x) = \int_{-\infty}^{+\infty}\ud x^{\prime}\, \mathcal{A}_F^{\Pi}(x,x^{\prime})\,\phi(x^{\prime})\, , \quad \phi \in L^2(\R,\ud x)\, ,
\end{equation}
with kernel
\begin{equation}
\label{AFker}
\mathcal{A}^{\Pi}_F(x,x^{\prime})= \frac{1}{2\pi}\int_{-\infty}^{+\infty}\ud q \, \hat F_p\left(q, x^{\prime}-x\right)\,\hat\Pi_p\left(x-x^{\prime},q-\frac{x+x^{\prime}}{2}\right)\,.
\end{equation}
\item In the above expression, $\hat F_p$ and $\hat\Pi_p$  stand respectively for the partial Fourier transforms of $F(q,p)$ and $\Pi(q,p)$  with respect to the $p$ variable
 \item Historical Weyl-Wigner ($\mathcal{W}-\mathcal{W}$) case holds for $\vap(z)= \Pi(q,p)=1$. Then 
\begin{equation}
\label{WWMM}
{\sf M}^{1}\equiv {\sf M}^{\mathcal{W}-\mathcal{W}}= 2{\sf P}\, , \quad \mathcal{A}^{\mathcal{W}-\mathcal{W}}_F(x,x^{\prime})= \frac{1}{\sqrt{2\pi}} \hat F_p\left(\frac{x+x^{\prime}}{2}, x^{\prime}-x\right)\,. 
\end{equation}
\item Another historical case: coherent state, or Berezin, or anti-Wick  quantization. From a normalised \textit{fiducial vector} $\eta(x)\in L^2(\R,\ud x)$, e.g. $\eta(x)= e_0(x)$, 
\begin{align}
\label{CSWHQ1}
{\sf M}^{|\eta\rg} &= |\eta\rg\lg\eta|\, , \quad \mathcal{M}^{|\eta\rg}(x,x^{\prime})=\overline{\eta(x^{\prime})}\eta(x)\, , \\  
\label{CSWHQ1} \mathcal{A}^{|\eta\rg}_F(x,x^{\prime})& = \frac{1}{\sqrt{2\pi}}\int_{-\infty}^{+\infty}\ud q \, \hat F_p\left(q, x^{\prime}-x\right)\,\overline{\eta(x^{\prime}-q)}\, \eta(x-q)\, . 
\end{align}
\item If applicable from \eqref{traceMom}, one has the trace formula
\begin{equation}
\label{traceAvapf}
\mathrm{tr}\left(A^{\vap}_f\right)=  \int_{\mathbb{C}} f(z) \,\frac{\ud^2 z}{\pi}\, . 
\end{equation}

\item Complex conjugaison covariance. From
 \begin{equation}
 \label{adjAf}
\left(A^{\vap}_{f}\right)^\dag =  A^{{\sf P}\overline{\vap}}_{\overline{f}} \, ,
\end{equation}
we have
 \begin{equation}
 \label{cococo}
 A^{\vap}_{\overline{f}} = \left(A^{\vap}_{f}\right)^\dag\, , \forall \,f   \ \iff \   \overline{\varpi(-z)}=\varpi(z)\,  \, \forall \,z\, .
\end{equation}
\item Translational covariance
\begin{equation}
\label{covquant}
A^{\vap}_{f(z-z_0)} = D(z_0) A^{\vap}_{f(z)} D(z_0)^\dag\, .
\end{equation}
\item Parity covariance
\begin{equation}
\label{quantvarpi2}
 A^{\vap}_{f(-z)} = {\sf P} A^{\vap}_{f(z)} {\sf P}\, , \, \forall \,f\  \  \iff \   \varpi(z)=\varpi(-z)\, , \,\forall \,z\, .
\end{equation}
\item Rotational covariance 
\begin{equation}
\label{rotcovAf}
U_{\mathbb{T}}(\theta)A^{\vap}_f U_{\mathbb{T}}(-\theta)= A^{\vap}_{T(\theta)f}  \ \iff \  \varpi\left(e^{\ii \theta}z\right)= \varpi(z) \, , \, \forall \,z\,, \theta  \ \iff \  \sf M^{\vap} \ \mbox{diagonal}\, , 
\end{equation}
where $T(\theta)f(z):= f\left(e^{-\ii \theta} z\right)$.
\item Fundamental separation formulae 
\begin{enumerate}
  \item[$\btr$] If $F(q,p)$ is a function of $q$ only, $F(q,p) \equiv u(q)$, then $A^{\Pi}_u$ depends on $Q$ only
  \begin{equation}
\label{Iquq}
A^{\Pi}_u= \frac{1}{\sqrt{2\pi}}\,u\ast \overline{\mathcal{F}}[\Pi(0,\cdot)](Q)\, ,
\end{equation}
where $ \overline{\mathcal{F}}$ is the inverse 1-D Fourier transform
\begin{equation}
\label{1D Fourier}
\mathcal{F}[h](x) =  \frac{1}{\sqrt{2\pi}}\,\int_{-\infty}^{+\infty} e^{-ikx}\,h(x)\,\ud x\, , \quad \overline{\mathcal{F}}[u](x) =  \frac{1}{\sqrt{2\pi}}\,\int_{-\infty}^{+\infty} e^{ikx}\,h(x)\,\ud x\, .
\end{equation}
  \item[$\btr$]  If $F(q,p)$ is a function of $p$ only, $F(q,p) \equiv v(p)$, then $A^{\Pi}_v$ depends on $P$ only
  \begin{equation}
\label{Iqvp}
A^{\Pi}_v= \frac{1}{\sqrt{2\pi}}\,v\ast \overline{\mathcal{F}}[\Pi(\cdot,0)](P)\, .
\end{equation}
\item[$\btr$] Similar formulae exist for holomorphic and anti-holomorphic functions $f(z)$
  \end{enumerate}
\end{itemize}
\section{Quantization of derivatives}
\label{quantderiv}
\begin{itemize}
  \item From \eqref{propdisp3}, \eqref{symFourder3} and \eqref{symFourder4} we derive easily\begin{equation}
\label{quantderf}
A^{\vap}_{\partial_{z}f}=-A^{\bar z\vap}_f= - [\adg, A^{\vap}_f] \,, \qquad A^{\vap}_{\partial_{\bar z}f}=  A^{z\vap}_f=  [a, A^{\vap}_f]\, , 
\end{equation}
\begin{equation}
\label{quantderfqp}
A^{\Pi}_{\partial_{q}F}=-A^{\ii p \Pi}_F=\ii\, [P, A^{\Pi}_F] \,, \qquad A^{\Pi}_{\partial_{p}F}=  A^{-\ii q\Pi}_F= -\ii \,[Q, A^{\Pi}_F]\, .
\end{equation}
  \item With Lie algebra adjoint action notation
  \begin{equation}
\label{defadjoint}
\mathrm{ad}_X(Y)= [X,Y]\,, 
\end{equation}  
we have more generally
\begin{equation}
\label{qdelidelif}
A^{\vap}_{\partial^{\bar i}_{z}\partial^i_{\bar z}f} = (-1)^{\bar i}\,\left(\mathrm{ad}^i_a\, \mathrm{ad}^{\bar i}_{\adg}\right)\left( A^{\vap}_f\right)\, ,
\end{equation} 
\begin{equation}
\label{qdelidelifqp}
A^{\Pi}_{\partial^{k}_{q}\partial^l_{p}F} = (\ii)^{k-l}\,\left(\mathrm{ad}^k_P\, \mathrm{ad}^{l}_{Q}\right)\left( A^{\Pi}_F\right)\, . 
\end{equation} 
 \item Here we recall the Jacobi identity
  \begin{equation}
\label{idJacobi}
[X,[Y,Z]] + [Z,[X,Y]]+ [Z,[X,Y]] = 0
\end{equation}
from which it results
\begin{equation}
\label{adadcom}
\left[\mathrm{ad}_X, \mathrm{ad}_Y\right](Z)= \mathrm{ad}_{[X,Y]}(Z)\, ,
\end{equation}
and so in our above particular case 
\begin{equation}
\label{comadad}
\left[\mathrm{ad}_a, \mathrm{ad}_{\adg}\right](Z) =  \mathrm{ad}_{I}(Z)= 0\, , \quad \left[\mathrm{ad}_Q, \mathrm{ad}_{P}\right](Z) =  \mathrm{ad}_{\ii I}(Z)= 0\,.
\end{equation}
\end{itemize}
\section{Quantization and  product(s)}
\label{quantprod}
\subsection{With $z$, $\bar z$ variables}
\begin{itemize}
  \item Quantization of product
  \begin{align}
\label{WHQproduct1}
A^{\vap}_{fg}&=\int_{\mathbb{C}} \int_{\mathbb{C}} \frac{\ud^2 z}{\pi} \frac{\ud^2 z^{\prime}}{\pi}\, e^{-\frac{1}{2}\,z\circ z^{\prime}} \, \varpi(z + z^{\prime})D(z)\, \omfs[f](z)\, D(z^{\prime})\, \omfs[g](z^{\prime}) \\
\label{WHQproduct2}&= \int_{\mathbb{C}} \int_{\mathbb{C}} \frac{\ud^2 z}{\pi} \frac{\ud^2 z^{\prime}}{\pi}\, e^{-\frac{1}{2}\,z\circ z^{\prime}} \, \varpi(z + z^{\prime})D(z)\, \omfs[g](z)\, D(z^{\prime})\, \omfs[f](z^{\prime}) \\
\label{WHQproduct3}&=\sum_{i,\bar i,j,\bar j} (-1)^{\bar i+ \bar j}\,a_{i\bar i j\bar j}\, A^{\vap}_{\partial^{i}_{\bar z}\partial^{\bar i}_{z}\,f} A^{\vap}_{\partial^{j}_{\bar z}\partial^{\bar j}_{z}\,g}= A^{\vap}_{f}A^{\vap}_{g}+ \dotsb\\
\label{WHQproduct4}&=\sum_{i,\bar i,j,\bar j} (-1)^{\bar i+\bar j}\,a_{i\bar i j\bar j}\, A^{\vap}_{\partial^{i}_{\bar z}\partial^{\bar i}_{z}\,g} A^{\vap}_{\partial^{j}_{\bar z}\partial^{\bar j}_{z}\,f}= A^{\vap}_{g}A^{\vap}_{f}+ \dotsb\, , \end{align}
where coefficients are defined by the (if making sense) development
\begin{equation}
\label{devarpi1}
e^{-\frac{1}{2}\,z\circ z^{\prime}} \, \frac{\varpi(z + z^{\prime})}{\vap(z)\,\vap(z^{\prime})}= \sum_{i,\bar i,j,\bar j}a_{i\bar i j\bar j}\, z^i\,\bar z^{\bar i}\, {z^{\prime}}^j\,\bar {z^{\prime}}^{\bar j}\, , \quad a_{0000}= 1\, .
\end{equation}
\item Note that the expression \eqref{WHQproduct3} provides a sort of inverse Moyal product on the level of operators
 \item Poisson bracket\\
 From 
 \begin{equation}
\label{poissonqpz}
\{ f,g\}:= \ii \,\left(\partial_{\bar z}f\,\partial_{z} g- \partial_{ z}f\,\partial_{\bar z}g\right)= \{F,G\}= \partial_q F\,\partial_p G-\partial_p F\,\partial_q G\, , 
\end{equation}
or, after restoring physical dimensions along \eqref{physdim} and \eqref{pdqpzbz},
 \begin{equation}
\label{poissonqpzpd}
\{f,g\}:= - \frac{1}{\ii \hbar}\,\left(\partial_{\bar z}f\,\partial_{z} g- \partial_{ z}f\,\partial_{\bar z}g\right)= \{F,G\}= \partial_q F\,\partial_p G-\partial_p F\,\partial_q G\, , 
\end{equation}
   \begin{align}
\label{WHQpoisson1}
A^{\vap}_{\{f,g\}}&=\frac{\ii}{\hbar}\,\sum_{i,\bar i,j,\bar j} (-1)^{\bar i+\bar j}\,a_{i\bar i j\bar j}\,\left[ A^{\vap}_{\partial^{i+1}_{\bar z}\partial^{\bar i}_{z}\,f} A^{\vap}_{\partial^{j}_{\bar z}\partial^{\bar j +1}_{z}\,g}-A^{\vap}_{\partial^{i}_{\bar z}\partial^{\bar i +1}_{z}\,f} A^{\vap}_{\partial^{j+1}_{\bar z}\partial^{\bar j }_{z}\,g}\right]\\
 \nonumber &= \frac{\ii}{\hbar}\,\sum_{i,\bar i,j,\bar j} a_{i\bar i j\bar j}\,\times\\
 \label{WHQpoisson2}& \times \left[\left(\mathrm{ad}^i_a\,\mathrm{ad}^{\bar i+1}_{\adg}\right)\left( A^{\vap}_f\right)\,\left(\mathrm{ad}^{j+1}_a\, \mathrm{ad}^{\bar j}_{\adg}\right)\left( A^{\vap}_g\right) - \left(\mathrm{ad}^{i+1}_a\, \mathrm{ad}^{\bar i}_{\adg}\right)\left( A^{\vap}_f\right)\,\left(\mathrm{ad}^j_a\, \mathrm{ad}^{\bar j+1}_{\adg}\right)\left( A^{\vap}_g\right) \right]\\
\label{WHQpoisson3}&= \frac{\ii}{\hbar}\,\left[ A^{\vap}_{\partial_{\bar z}\,f} A^{\vap}_{\partial_{z}\,g}-A^{\vap}_{\partial_{z}\,f} A^{\vap}_{\partial_{\bar z}\,g}\right]+ \dotsb
\end{align}
 \item Product of quantizations
 \begin{equation}
\label{WHQproduct}
A^{\vap}_{f}\,A^{\vap}_g= A^{\vap}_{f\star_{\varpi} g}\, ,
\end{equation}
where the ``$\vap$-Moyal product'' is defined by
\begin{equation}
\label{vapmoyal}
(f\star_{\varpi} g) (z)=  \sum_{i,\bar i,j,\bar j}(-1)^{\bar i+ \bar j}\,\tilde{a}_{i\bar i j\bar j}\, \left(\partial^{i}_{\bar z}\partial^{\bar i}_{z}\,f\right)\,\left(\partial^{j}_{\bar z}\partial^{\bar j}_{z}\,g\right)\, ,
\end{equation} 
with coefficients defined by the (if making sense) expansion
\begin{equation}
\label{devarpi2}
e^{\frac{1}{2}\,z\circ z^{\prime}} \, \frac{\vap(z)\,\vap(z^{\prime})}{\varpi(z + z^{\prime})}= \sum_{i,\bar i,j,\bar j}\tilde{a}_{i\bar i j\bar j}\, z^i\,\bar z^{\bar i}\, {z^{\prime}}^j\,\bar {z^{\prime}}^{\bar j}\, , \quad \quad \tilde{a}_{0000}= 1\, .
\end{equation}
Since coefficients $\tilde{a}_{i\bar i j\bar j}$ result from the inverse of \eqref{devarpi1}, we have the convolution relations with coefficients $a_{i\bar i j\bar j}$
\begin{equation}
\label{convab}
\sum_{i,\bar i,j,\bar j}a_{i\bar i j\bar j}\,\tilde{a}_{k-i\,\bar k-\bar i \,l-  j\,\bar l-\bar j} = 0
\end{equation}
for all $k$, $\bar k$, $l$, $\bar l$, such that $ k + \bar k + l + \bar l \geq 1$.
\item Commutator of quantizations
 \begin{align}
\nonumber\left[A^{\vap}_{f}\,,\,A^{\vap}_g\right]&= A^{\vap}_{f\star_{\varpi} g - g\star_{\varpi} f }\\
\label{comquant1}&= \sum_{i,\bar i,j,\bar j}(-1)^{\bar i+ \bar j}\,\tilde{a}_{i\bar i j\bar j}\, \left[A^{\vap}_{\left(\partial^{i}_{\bar z}\partial^{\bar i}_{z}\,f\right)\,\left(\partial^{j}_{\bar z}\partial^{\bar j}_{z}\,g\right)}-A^{\vap}_{\left(\partial^{i}_{\bar z}\partial^{\bar i}_{z}\,g\right)\,\left(\partial^{j}_{\bar z}\partial^{\bar j}_{z}\,f\right)}\right]\\
\label{comquant2}&= i\,A^{\vap}_{\{f,g\}} + \dotsb
\end{align}
\end{itemize}

\subsection{With $q$, $p$ variables}
\begin{itemize}
  \item Quantization of product
  \begin{align}
\label{WHQprodqp1}
\nonumber A^{\Pi}_{FG}&=\int_{\R^2} \int_{\R^2} \frac{\ud q \,\ud p}{2\pi} \frac{\ud q^{\prime}\,\ud p^{\prime}}{2\pi}\, e^{\frac{\ii}{2}\,(q\,p^{\prime}-p\,q^{\prime})} \, \Pi(q + q^{\prime}, p + p^{\prime})\times\\&\times\mathcal{D}(q,p)\, \omFs[F](q,p)\, \mathcal{D}(q^{\prime},p^{\prime})\, \omFs[G](q^{\prime},q^{\prime}) \\
\label{WHQprodqp2}&=\sum_{k,l,m,n} \ii^{k-l+m-n}\,\Da_{k l m n }\, A^{\Pi}_{\partial^{l}_{q}\partial^{k}_{p}\,F} A^{\Pi}_{\partial^{n}_{q}\partial^{m}_{p}\,G}= A^{\Pi}_{F}A^{\Pi}_{G}+ \dotsb\, , \end{align}
where coefficients are defined by the (if making sense) development
\begin{equation}
\label{devPi1}
e^{\frac{\ii}{2}\,(q\,p^{\prime}-p\,q^{\prime})} \, \frac{\Pi(q + q^{\prime}, p + p^{\prime})}{\Pi(q , p )\,\Pi(q^{\prime},p^{\prime})}= \sum_{k,l,k^{\prime}l^{\prime}}\Da_{k l k^{\prime}l^{\prime}}\, q^k\,p^{l}\, {q^{\prime}}^{k^{\prime}}\,{p^{\prime}}^{l^{\prime}}\, , \quad \Da_{0000}= 1\, .
\end{equation}
 \item Poisson bracket\\
 From 
 \begin{equation}
\label{poissonqp}
 \{F,G\}= \partial_q F\,\partial_p G-\partial_p F\,\partial_q G\, , 
\end{equation}
   \begin{align}
\label{WHQpoissonqp1}
A^{\Pi}_{\{F,G\}}&=\sum_{k,l,m,n} \ii^{k-l+m-n}\,\Da_{k l m n }\,\left[ A^{\Pi}_{\partial^{l+1}_{q}\partial^{k}_{p}\,F} A^{\Pi}_{\partial^{n}_{q}\partial^{m+1}_{p}\,G} -A^{\Pi}_{\partial^{l}_{q}\partial^{k+1}_{p}\,F} A^{\Pi}_{\partial^{n+1}_{q}\partial^{m}_{p}\,G}\right]\\
 \nonumber &= \sum_{k,l,m,n} \Da_{k lmn}\,\times\\
 \label{WHQpoissonqp2}& \times \left[\left(\mathrm{ad}^k_Q\,\mathrm{ad}^{l+1}_{P}\right)\left( A^{\Pi}_F\right)\,\left(\mathrm{ad}^{m+1}_Q\, \mathrm{ad}^{n}_{P}\right)\left( A^{\Pi}_G\right) - \left(\mathrm{ad}^{k+1}_Q\, \mathrm{ad}^{l}_{P}\right)\left( A^{\Pi}_F\right)\,\left(\mathrm{ad}^m_Q\, \mathrm{ad}^{n+1}_{P}\right)\left( A^{\Pi}_G\right) \right]\\
\label{WHQpoissonqp3}&= \left[ A^{\Pi}_{\partial_{q}\,F} A^{\Pi}_{\partial_{p}\,G}-A^{\Pi}_{\partial_{p}\,F} A^{\Pi}_{\partial_{q}\,G}\right]+ \dotsb
\end{align}
 \item Product of quantizations
 \begin{equation}
\label{WHQprodqp}
A^{\Pi}_{F}\,A^{\Pi}_G= A^{\Pi}_{F\star_{\Pi} G}\, ,
\end{equation}
where the ``$\Pi$-Moyal product'' is defined by
\begin{equation}
\label{Pimoyalqp}
(F\star_{\varpi} G) (q,p)=  \sum_{k,l,m,n}\ii^{k+m-l-n}\,\TDa_{klmn}\, \left(\partial^{l}_{q}\partial^{k}_{p}\,F\right)\,\left(\partial^{n}_{q}\partial^{m}_{p}\,G\right)\, ,
\end{equation} 
with coefficients defined by the (if making sense) expansion
\begin{equation}
\label{devPi2}
e^{-\frac{\ii}{2}\,(q\,p^{\prime}-p\,q^{\prime})} \, \frac{\Pi(q , p )\,\Pi(q^{\prime},p^{\prime})}{\Pi(q + q^{\prime}, p + p^{\prime})}= \sum_{k,l,k^{\prime}l^{\prime}}\TDa_{k l k^{\prime}l^{\prime}}\, q^k\,p^{l}\, {q^{\prime}}^{k^{\prime}}\,{p^{\prime}}^{l^{\prime}}\, , \quad \TDa_{0000}= 1\, .
\end{equation}
Since coefficients $\TDa_{k l k^{\prime}l^{\prime}}$ result from the inverse of \eqref{devPi1}, we have the convolution relations with coefficients $\Da_{k l k^{\prime}l^{\prime}}$
\begin{equation}
\label{convab}
\sum_{m,n,m^{\prime},n^{\prime}}\Da_{m,n,m^{\prime},n^{\prime}}\,\TDa_{k-m,l-n,k^{\prime}-m^{\prime},l^{\prime}-n^{\prime}} = 0
\end{equation}
for all $k$, $l$, $k^{\prime}$, $l^{\prime}$, such that $ k  +l + k^{\prime}+l^{\prime}\geq 1$.
\item Commutator of quantizations
 \begin{align}
\nonumber\left[A^{\Pi}_{F}\,,\,A^{\Pi}_G\right]&= A^{\Pi}_{F\star_{\Pi} G - G\star_{\Pi} F}\\
\label{comquant1qp}&=  \sum_{k,l,m,n}\ii^{k+m-l-n}\,\TDa_{klmn}\,\left[ A^{\Pi}_{\left(\partial^{l}_{q}\partial^{k}_{p}\,F\right)\,\left(\partial^{n}_{q}\partial^{m}_{p}\,G\right)}- A^{\Pi}_{\left(\partial^{l}_{q}\partial^{k}_{p}\,G\right)\,\left(\partial^{n}_{q}\partial^{m}_{p}\,F\right)}\right]\\
\label{comquant2qp}&= i\,A^{\Pi}_{\{F,G\}} + \dotsb
\end{align}
\end{itemize}
 
\section{Interlude I: expressions for expansion coefficients, variables $z$, $\bar z$, arbitrary $\vap$}
\label{expcoefI}
\begin{itemize}
  \item Supposing real analyticity for weight function $\vap$, we define expansion coefficients for it by
  \begin{equation}
\label{expvap}
\vap(z)= \sum_{i,\bar i} c_{i\bar i}\, z^i\,\bar z^{\bar i}\, , \quad c_{00}= 1\, , \quad  c_{i\bar i} = \left.\frac{1}{i!\bar i!}\,\partial_{z}^{i}\,\partial_{\bar z}^{\bar i}\,\vap(z)\right\vert_{z=0}\, ,
\end{equation}
and for its inverse
  \begin{equation}
\label{expivap}
\frac{1}{\vap(z)}= \sum_{i,\bar i} \tilde{c}_{i\bar i}\, z^i\,\bar z^{\bar i}\, , \quad \tilde{c}_{00}= 1\, , \quad  \tilde{c}_{i\bar i} = \left.\frac{1}{i!\bar i!}\,\partial_{z}^{i}\,\partial_{\bar z}^{\bar i}\,\frac{1}{\vap(z)}\right\vert_{z=0}\, ,
\end{equation}
with the discrete convolution equation holding for all $n$, $\bar n$ such that $n+\bar n \geq 1$,
\begin{equation}
\label{relctc}
\sum_{i , \bar i } c_{n-i, \bar n-\bar i}\,\tilde{c}_{i\bar i}= 0\,. 
\end{equation}
\item For instance
\begin{align}
\label{ctcr01}
  \tilde{c}_{10}  &=  - c_{10}\, , \quad  \tilde{c}_{01} = - c_{01}\,,\\
  \label{ctcr11}   \tilde{c}_{11}  &=  - c_{11} + 2c_{10}\,c_{01}\,,\\
   \label{ctcr20}   \tilde{c}_{20}  &=  - c_{20} + c^2_{10}\,,\quad  \tilde{c}_{02}  =  - c_{02} + c^2_{01}\,. 
\end{align}
\item There results expansion coefficients for the ratios 
 \begin{align}
\label{expratSP1}
\frac{\varpi(z + z^{\prime})}{\vap(z)\,\vap(z^{\prime})}&= \sum_{i,\bar i,j,\bar j}d_{i\bar i j\bar j}\, z^i\,\bar z^{\bar i}\, {z^{\prime}}^j\,\bar {z^{\prime}}^{\bar j}\,,  \quad  d_{0000}= 1\, , \\
\label{expratSP1} d_{i\bar ij\bar j} &= \left.\frac{1}{i!\bar i!j!\bar j!}\,\partial_{z}^{i}\,\partial_{\bar z}^{\bar i}\,\,\partial_{z^{\prime}}^{j}\,\partial_{\bar z^{\prime}}^{\bar j}\,\frac{\varpi(z + z^{\prime})}{\vap(z)\,\vap(z^{\prime})}\right\vert_{z,z^{\prime}=0}\, .
\end{align}
 \begin{align}
\label{expratPS1}
\frac{\vap(z)\,\vap(z^{\prime})}{\varpi(z + z^{\prime})}&= \sum_{i,\bar i,j,\bar j}\tilde{d}_{i\bar i j\bar j}\, z^i\,\bar z^{\bar i}\, {z^{\prime}}^j\,\bar {z^{\prime}}^{\bar j}\,, \quad \tilde{d}_{0000}= 1\, , \\
\label{expratSP2} \tilde{d}_{i\bar ij\bar j} &= \left.\frac{1}{i!\bar i!j!\bar j!}\,\partial_{z}^{i}\,\partial_{\bar z}^{\bar i}\,\,\partial_{z^{\prime}}^{j}\,\partial_{\bar z^{\prime}}^{\bar j}\,\frac{\vap(z)\,\vap(z^{\prime})}{\varpi(z + z^{\prime})}\right\vert_{z,z^{\prime}=0}
\, .
\end{align}
with the discrete convolution equation  holding for all $k$, $\bar k$, $l$, $\bar l$ such that $k+\bar k + l+\bar l\geq 1$,
\begin{equation}
\label{relctc}
\sum_{i , \bar i , j, \bar j} d_{k-i, \bar k-\bar i, l -j, \bar l -\bar j}\,\tilde{d}_{i\bar i j \bar j}= 0\,. 
\end{equation}
\item Expressions of $d_{i\bar i j\bar j}$ and $\tilde{d}_{i\bar i j\bar j}$ in terms of $c_{i\bar i}$ and $\tilde c_{i\bar i}$
\begin{align}
\label{dctc}
  d_{i\bar i j\bar j}  &=  \sum_{s,\bar s, r,\bar r} \binom{i+j-s}{i-r} \, \binom{\bar i + \bar j - \bar s}{\bar i - \bar r}\, c_{i+j-s, \bar i + \bar j - \bar s}\, \tilde{c}_{s-r, \bar s-\bar r}\, \tilde{c}_{r \bar r}\, , \\
 \label{tdctc}  \tilde{d}_{i\bar i j\bar j} & =  \sum_{s,\bar s, r,\bar r} \binom{i+j-s}{i-r} \, \binom{\bar i + \bar j - \bar s}{\bar i - \bar r}\, \tilde{c}_{i+j-s, \bar i + \bar j - \bar s}\, c_{s-r, \bar s-\bar r}\, c_{r \bar r}\, .
\end{align}
\item Properties of coefficients $d_{i\bar i j\bar j}$ and $\tilde{d}_{i\bar i j\bar j}$
\begin{equation}
\label{propdibijbj}
d_{0000}= 1\, , \quad d_{i\bar i00}= 0 =   d_{00j\bar j}\ \forall\, i\, , \, \bar i \, ,\, j\, , \bar j\, , \,   i+\bar i \geq 1\,, \, j+\bar j\geq 1\, . 
\end{equation}
\begin{equation}
\label{proptdibijbj}
\tilde{d}_{0000}= 1\, , \quad \tilde{d}_{i\bar i00}= 0 =   \tilde{d}_{00j\bar j}\ \forall\, i\, , \, \bar i \, ,\, j\, , \bar j\, , \,   i+\bar i \geq 1\,, \, j+\bar j\geq 1\, . 
\end{equation}
\item Other particular expressions
\begin{equation}
\label{d0ij0} d_{i 00\bar j}  =  \sum_{s,\bar s}  c_{i-s,  \bar j - \bar s}\, \tilde{c}_{s 0}\, \tilde{c}_{0 \bar s}\, , \qquad 
d_{0\bar i j 0}  =  \sum_{s,\bar s}  c_{j-s,  \bar i- \bar s}\, \tilde{c}_{s 0}\, \tilde{c}_{0 \bar s}\,.
\end{equation}
 \item Coefficients $d_{i\bar i j\bar j}$ for $i+\bar i+ j +\bar j \leq 2$ are given by
  \begin{align}
\label{dij1}
   d_{0000} & = 1\,, \  d_{1000} =d_{0100} =d_{0010} =d_{0001} =0\, ,  \\
 \label{dij2}  d_{1001}  &= c_{11} - c_{10}\,c_{01} = d_{0110} \,,\\
 \label{dij3}  d_{1010}  &= 2c_{20} - c^2_{10}\,, \quad  d_{0101}= 2c_{02} -c^2_{01} \, , \\
\label{dij4}  d_{2000} &=d_{0200} =d_{0020} =d_{0002} =0 \, ,\\
\label{dij5}  d_{1100} & =d_{0011} =0 \, . 
  \end{align}
\item Expressions of $a_{i\bar i j\bar j}$ and $\tilde{a}_{i\bar i j\bar j}$ in terms of $d_{i\bar i j\bar j}$ and $\tilde{d}_{i\bar i j\bar j}$ respectively
\begin{align}
\label{areld}
  a_{i\bar i j\bar j}  &=  \sum_{ k=0}^{\min(i,\bar j)}\sum_{ l=0}^{\min(\bar i,j)} \frac{(-1)^k}{k!\,l!}\,\frac{1}{2^{k+l}}\, d_{i-k,\bar i - l, j-l,\bar j-k }\,  , \\
 \label{tareltd}  \tilde{a}_{i\bar i j\bar j} & = =  \sum_{ k=0}^{\min(i,\bar j)}\sum_{ l=0}^{\min(\bar i,j)} \frac{(-1)^l}{k!\,l!}\,\frac{1}{2^{k+l}}\, \tilde{d}_{i-k,\bar i - l, j-l,\bar j-k }\,  . 
  \end{align}
  \item In particular
  \begin{equation}
\label{ i0j0} a_{i 0 j 0} = d_{i 0j 0}\, ,\quad a_{0\bar i 0\bar j} = d_{0\bar i 0\bar j}\, . 
\end{equation}
  \item Properties of coefficients $a_{i\bar i j\bar j}$ and $\tilde{a}_{i\bar i j\bar j}$
\begin{equation}
\label{propaibijbj}
a_{0000}= 1\, , \quad a_{i\bar i00}= 0 =   a_{00j\bar j}\ \forall\, i\, , \, \bar i \, ,\, j\, , \bar j\, , \,   i+\bar i \geq 1\,, \, j+\bar j\geq 1\, . 
\end{equation}
\begin{equation}
\label{proptaibijbj}
\tilde{a}_{0000}= 1\, , \quad \tilde{a}_{i\bar i00}= 0 =   \tilde{a}_{00j\bar j}\ \forall\, i\, , \, \bar i \, ,\, j\, , \bar j\, , \,   i+\bar i \geq 1\,, \, j+\bar j\geq 1\, . 
\end{equation}
  \item Coefficients $a_{i\bar i j\bar j}$ for $i+\bar i+ j +\bar j \leq 2$ are given by
  \begin{align}
\label{aij1}
   a_{0000} & = 1\,, \  a_{1000} =a_{0100} =a_{0010} =a_{0001} =0\, ,  \\
 \label{aij2}  a_{1001}  &= c_{11} - c_{10}\,c_{01} - \frac{1}{2}\,, \quad a_{0110} = c_{11} - c_{10}\,c_{01} + \frac{1}{2}\,,\\
 \label{aij3}  a_{1010}  &= 2c_{20} - c^2_{10}\,, \quad  a_{0101}= 2c_{02} -c^2_{01} \, , \\
\label{aij4}  a_{2000} &=a_{0200} =a_{0020} =a_{0002} =0 \, .  
  \end{align}
  \item Coefficients $\tilde{a}_{i\bar i j\bar j}$ for $i +\bar i + j +\bar j \leq 2$ are given by
   \begin{align}
\label{bij1}
   \tilde{a}_{0000} & = 1\,,\  \tilde{a}_{1000} =\tilde{a}_{0100} =\tilde{a}_{0010} =\tilde{a}_{0001} =0\, ,  \\
 \label{bij2}  \tilde{a}_{1001}  &= -c_{11} + c_{10}\,c_{01} + \frac{1}{2}= - a_{1001} \,, \quad  \tilde{a}_{0110} = -c_{11} + c_{10}\,c_{01} - \frac{1}{2} = - a_{0110}\,,\\
 \label{bij3}  \tilde{a}_{1010}  &= - 2c_{20}= - a_{1010} -c^2_{10} \,, \quad  \tilde{a}_{0101}= -2c_{02} = - a_{0101} -c^2_{01} \, , \\
\label{bij4}  \tilde{a}_{2000} &=\tilde{a}_{0200} =\tilde{a}_{0020} =\tilde{a}_{0002} =0 \, . 
  \end{align}
\end{itemize}

\section{Interlude II: expressions for expansion coefficients, variables $q$, $ p$, arbitrary $\Pi$}
\label{expcoefII}
\begin{itemize}
  \item Supposing analyticity for weight function $\Pi$, we define expansion coefficients for it by
  \begin{equation}
\label{expPi}
\Pi(q,p)= \sum_{k,l} \Dc_{kl}\, q^k\,p^{l}\, , \quad \Dc_{00}= 1\, , \quad  \Dc_{kl} = \left.\frac{1}{k! l!}\,\partial_{q}^{k}\,\partial_{p}^{l}\,\Pi(q,p)\right\vert_{q=0=p}\, ,
\end{equation}
and for its inverse
  \begin{equation}
\label{expiPi}
\frac{1}{\Pi(q,p)}= \sum_{k,l} \TDc_{kl}\, q^k\,p^{\bar i}\, , \quad \TDc_{00}= 1\, , \quad  \TDc_{kl} = \left.\frac{1}{k!l!}\,\partial_{q}^{k}\,\partial_{p}^{l}\,\frac{1}{\Pi(q,p)}\right\vert_{q=0=p}\, ,
\end{equation}
with the discrete convolution equation holding for all $m$, $n$ such that $m+n \geq 1$,
\begin{equation}
\label{relCTC}
\sum_{k, l } \Dc_{m-k, n-l}\,\TDc_{kl}= 0\,. 
\end{equation}
\item For instance
\begin{align}
\label{CTCr01}
  \TDc_{10}  &=  - \Dc_{10}\, , \quad  \TDc_{01} = - \Dc_{01}\,,\\
  \label{CTC11}   \TDc_{11}  &=  - \Dc_{11} + 2\Dc_{10}\,\Dc_{01}\,,\\
   \label{CTC20}   \TDc_{20}  &=  - \Dc_{20} + \Dc^2_{10}\,,\quad  \TDc_{02}  =  - \Dc_{02} + \Dc^2_{01}\,. 
\end{align}
\item There results expansion coefficients for the ratios 
 \begin{align}
\label{expratSP1qp}
\frac{\Pi(q + q^{\prime},p+ p^{\prime})}{\Pi(q,p)\,\Pi(q^{\prime},p^{\prime})}&= \sum_{k,l,k^{\prime},l^{\prime}}\Dd_{klk^{\prime}l^{\prime}}\, q^k\,p^{l}\, {q^{\prime}}^{k^{\prime}}\, {p^{\prime}}^{l^{\prime}}\,,  \quad  \Dd_{0000}= 1\, , \\
\label{expratSP1qp} \Dd_{klk^{\prime}l^{\prime}} &= \left.\frac{1}{k!l!k^{\prime}!l^{\prime}!}\,\partial_{q}^{k}\,\partial_{p}^{l}\,\,\partial_{q^{\prime}}^{k^{\prime}}\,\partial_{\bar p^{\prime}}^{l^{\prime}}\,\frac{\Pi(q + q^{\prime}, p+p^{\prime})}{\Pi(q,p)\,\Pi(q^{\prime},p^{\prime})}\right\vert_{q,p,q^{\prime},p^{\prime}=0}\, .
\end{align}
 \begin{align}
\label{expratPS1qp}
\frac{\Pi(q,p)\,\Pi(q^{\prime},p^{\prime})}{\Pi(q + q^{\prime},p+ p^{\prime})}&= \sum_{k,l,k^{\prime},l^{\prime}}\TDd_{klk^{\prime}l^{\prime}}\, q^k\,p^{l}\, {q^{\prime}}^{k^{\prime}}\, {p^{\prime}}^{l^{\prime}}\,,  \quad  \TDd_{0000}= 1\, , \\
\label{expratSP2qp} \TDd_{klk^{\prime}l^{\prime}} &= \left.\frac{1}{k!l!k^{\prime}!l^{\prime}!}\,\partial_{q}^{k}\,\partial_{p}^{l}\,\,\partial_{q^{\prime}}^{k^{\prime}}\,\partial_{ p^{\prime}}^{l^{\prime}}\,\frac{\Pi(q,p)\,\Pi(q^{\prime},p^{\prime})}{\Pi(q + q^{\prime},p+ p^{\prime})}\right\vert_{q,p,q^{\prime},p^{\prime}=0}\, .
\end{align}
with the discrete convolution equation  holding for all $k$, $l$, $k^{\prime}$, $l^{\prime}$ such that $k+ l+ k^{\prime}+ l^{\prime}\geq 1$,
\begin{equation}
\label{relctcqp}
\sum_{m,n,m^{\prime}, n^{\prime}} \Dd_{k-m, l-\bar i, l -n, k^{\prime} -m^{\prime}, l^{\prime}-n^{\prime}}\,\TDd_{mnm^{\prime}n^{\prime}}= 0\,. 
\end{equation}
\item Expressions of $\Dd_{klk^{\prime}l^{\prime}}$ and $\TDd_{klk^{\prime}l^{\prime}}$ in terms of $\Dc_{kl}$ and $\TDc_{kl}$
\begin{align}
\label{dctcqp}
\Dd_{klk^{\prime}l^{\prime}}  &=  \sum_{r,s,m,n} \binom{r}{m} \, \binom{s}{n}\, \Dc_{rs}\, \TDc_{k-r+m,l-s+n }\, \TDc_{k^{\prime}-m,l^{\prime}-n}\, , \\
 \label{tdctcqp}  \TDd_{klk^{\prime}l^{\prime}} & =  \sum_{r,s,m,n} \binom{r}{m} \, \binom{s}{n}\, \TDc_{rs}\, \Dc_{k-r+m,l-s+n }\, \Dc_{k^{\prime}-m,l^{\prime}-n} \, .
\end{align}
\item Properties of coefficients $\Dd_{klk^{\prime}l^{\prime}}$ and $\TDd_{klk^{\prime}l^{\prime}}$
\begin{equation}
\label{propdklqp}
\Dd_{0000}= 1\, , \quad \Dd_{kl00}= 0 =   \Dd_{00k^{\prime}l^{\prime}}\ \forall\, k\, , \, l \, ,\, k^{\prime}\, , l^{\prime}\, , \,   k+l\geq 1\,, \, k^{\prime}+ l^{\prime}\geq 1\, . 
\end{equation}
\begin{equation}
\label{proptdklqp}
\TDd_{0000}= 1\, , \quad \TDd_{kl00}= 0 =   \TDd_{00k^{\prime}l^{\prime}}\ \forall\, k\, , \, l \, ,\, k^{\prime}\, , l^{\prime}\, , \,   k+l\geq 1\,, \, k^{\prime}+ l^{\prime}\geq 1 \, . 
\end{equation}
\item Other particular expressions
\begin{equation}
\label{d0lk0qp} \Dd_{k00l^{\prime}}  =  \sum_{r,s}  \Dc_{r s}\, \TDc_{k-r, 0}\, \TDc_{0, l^{\prime}-s}\, , \qquad 
\Dd_{0lk^{\prime} 0}  =  \sum_{r,s}  \Dc_{r s}\, \TDc_{0,l-s }\, \TDc_{k^{\prime}-r, 0}\,.
\end{equation}
\begin{equation}
\label{dk0k0qp} \Dd_{k0k^{\prime}0}  =  \sum_{r,m} \binom{r}{m}\, \Dc_{r 0}\, \TDc_{k-r+m, 0}\, \TDc_{ k^{\prime}-m,0}\, , \quad 
\Dd_{0l0l^{\prime}}  =  \sum_{s,n}  \binom{s}{n}\, \Dc_{0 s}\, \TDc_{0,l-s +n }\, \TDc_{0,l^{\prime}-n}\,.
\end{equation}
 \item Coefficients $\Dd_{klk^{\prime}l^{\prime}}$ for $k+l+k^{\prime}+l^{\prime}\leq 2$ are given by
  \begin{align}
\label{dkl1qp}
   \Dd_{0000} & = 1\,, \  \Dd_{1000} =\Dd_{0100} =\Dd_{0010} =\Dd_{0001} =0\, ,  \\
 \label{dkl2qp}  \Dd_{1001}  &= \Dc_{11} - \Dc_{10}\,\Dc_{01} = \Dd_{0110} \,,\\
 \label{dkl3qp}  \Dd_{1010}  &= 2\Dc_{20} - \Dc^2_{10}\,, \quad  \Dd_{0101}= 2\Dc_{02} -\Dc^2_{01} \, , \\
\label{dkl4qp}  \Dd_{2000} &=\Dd_{0200} =\Dd_{0020} =\Dd_{0002} =0 \, ,\\
\label{dkl5qp}  \Dd_{1100} & =\Dd_{0011} =0 \, . 
  \end{align}
\item Expressions of $\Da_{klk^{\prime}l^{\prime}}$ and $\TDa_{klk^{\prime}l^{\prime}}$ in terms of $\Dd_{klk^{\prime}l^{\prime}}$ and $\TDd_{klk^{\prime}l^{\prime}}$ respectively
\begin{align}
\label{areldqp}
 \Da_{klk^{\prime}l^{\prime}} &=  \sum_{ m=0}^{\min(k,l^{\prime})}\sum_{ n=0}^{\min(l,k^{\prime})} \frac{\ii^{m-n}}{m!\,n!}\,\frac{1}{2^{m+n}}\, \Dd_{k-m,l-n, k^{\prime}-n, l^{\prime}-m }\,  , \\
 \label{tareltdqp}   \TDa_{klk^{\prime}l^{\prime}} &=  \sum_{ m=0}^{\min(k,l^{\prime})}\sum_{ n=0}^{\min(l,k^{\prime})} \frac{\ii^{n-m}}{m!\,n!}\,\frac{1}{2^{m+n}}\, \TDd_{k-m,l-n, k^{\prime}-n, l^{\prime}-m }\,  . 
  \end{align}
  \item In particular
  \begin{equation}
\label{k0k0qp} \Da_{k0k^{\prime}0} = \Dd_{k0k^{\prime}0}\, ,\quad \Da_{0l0l^{\prime}} = \Dd_{0l0 l^{\prime}}\, . 
\end{equation}
  \item Properties of coefficients $\Da_{klk^{\prime}l^{\prime}}$ and $\TDa_{klk^{\prime}l^{\prime}}$
\begin{equation}
\label{propaklklqp}
\Da_{0000}= 1\, , \quad \Da_{kl00}= 0 =   \Da_{00k^{\prime}l^{\prime}}\ \forall\, k\, , \, l \, ,\, k^{\prime}\, , l^{\prime}\, , \,   k+l\geq 1\,, \, k^{\prime}+ l^{\prime}\geq 1\, . 
\end{equation}
\begin{equation}
\label{proptaklklqp}
\TDa_{0000}= 1\, , \quad \TDa_{kl00}= 0 =   \TDa_{00k^{\prime}l^{\prime}}\ \forall\, k\, , \, l \, ,\, k^{\prime}\, , l^{\prime}\, , \,   k+l\geq 1\,, \, k^{\prime}+ l^{\prime}\geq 1\, . 
\end{equation}
  \item Coefficients $\Da_{klk^{\prime}l^{\prime}}$ for $k+l+k^{\prime}+l^{\prime}\leq 2$ are given by
  \begin{align}
\label{akl1qp}
   \Da_{0000} & = 1\,, \  \Da_{1000} =\Da_{0100} =\Da_{0010} =\Da_{0001} =0\, ,  \\
 \label{akl2qp}  \Da_{1001}  &= \Dc_{11} - \Dc_{10}\,\Dc_{01} + \frac{\ii}{2}\,, \quad \Da_{0110} = \Dc_{11} - \Dc_{10}\,\Dc_{01} - \frac{\ii}{2}\,,\\
 \label{akl3qp}  \Da_{1010}  &= 2\Dc_{20} - \Dc^2_{10}\,, \quad  \Da_{0101}= 2\Dc_{02} -\Dc^2_{01} \, , \\
\label{akl4qp}  \Da_{2000} &=\Da_{0200} =\Da_{0020} =\Da_{0002} =0 \, .  
  \end{align}
\end{itemize}

\section{Interlude III: quantization results with Cahill-Glauber weight}
\label{expcoefIII}
\begin{itemize}
  \item Cahill-Glauber weight: $\vap(z) = e^{\frac{s}{2}\vert z \vert^2}$
  \item Coefficients are
  \begin{align}
\label{CGcoefa}
   a_{i\bar i j \bar j} &= \frac{1}{i!\,j!} \, \delta_{i\bar j}\, \delta_{\bar i j} \, \left(\frac{s-1}{2}\right)^i\, \left(\frac{s+1}{2}\right)^j\, , \\
 \label{CGcoefb}   \tilde{a}_{i\bar i j \bar j} & = (-1)^{i+j}a_{i\bar i j \bar j} \, . 
\end{align}
  \item Quantization of Poisson bracket 
   \begin{align}
\label{WHQpoissonCG1}
A^{\vap}_{\{f,g\}}&=\frac{\ii}{\hbar}\,\sum_{i,j} \frac{(-1)^{i+j}}{i!j!}\, \left(\frac{s-1}{2}\right)^i\, \left(\frac{s+1}{2}\right)^j\,\left[ A^{\vap}_{\partial^{i+1}_{\bar z}\partial^{j}_{z}\,f} A^{\vap}_{\partial^{j}_{\bar z}\partial^{i +1}_{z}\,g}-A^{\vap}_{\partial^{i}_{\bar z}\partial^{j +1}_{z}\,f} A^{\vap}_{\partial^{j+1}_{\bar z}\partial^{i }_{z}\,g}\right]\\
\nonumber &= \frac{\ii}{\hbar}\,\sum_{i,j} \frac{1}{i!j!}\, \left(\frac{s-1}{2}\right)^i\, \left(\frac{s+1}{2}\right)^j\times \\
 \label{WHQpoissonCG2} &\times \left[\left(\mathrm{ad}^{i}_a\, \mathrm{ad}^{j+1}_{\adg}\right)\left( A^{\vap}_f\right)\,\left(\mathrm{ad}^{j+1}_a\, \mathrm{ad}^{i}_{\adg}\right)\left( A^{\vap}_g\right) - \left(\mathrm{ad}^{i+1}_a\, \mathrm{ad}^{j}_{\adg}\right)\left( A^{\vap}_f\right)\,\left(\mathrm{ad}^j_a\, \mathrm{ad}^{i+1}_{\adg}\right)\left( A^{\vap}_g\right) \right]
\end{align}
\item $\vap$-Moyal product
\begin{equation}
\label{vapmoyalCG}
(f\star_{\varpi} g) (z)=  \sum_{i,j}\frac{1}{i!\,j!}  \, \left(\frac{s-1}{2}\right)^i\, \left(\frac{s+1}{2}\right)^j\, \left(\partial^{i}_{\bar z}\partial^{j}_{z}\,f\right)\,\left(\partial^{j}_{\bar z}\partial^{i}_{z}\,g\right)\, ,
\end{equation}
\item $\vap$-Moyal  commutator
\begin{align}
\nonumber (f\star_{\varpi} g) (z)- (g\star_{\varpi} f) (z)&= \sum_{i,j}\frac{1}{i!\,j!}  \, \left(\frac{s-1}{2}\right)^i\, \left(\frac{s+1}{2}\right)^j\times \\
\label{vapmoyalcomCG1}
&\times \left[\left(\partial^{i}_{\bar z}\partial^{j}_{z}\,f\right)\,\left(\partial^{j}_{\bar z}\partial^{i}_{z}\,g\right)-\left(\partial^{i}_{\bar z}\partial^{j}_{z}\,g\right)\,\left(\partial^{j}_{\bar z}\partial^{i}_{z}\,f\right)\right]\\
\nonumber  &=\sum_{i,j}\frac{\mathrm{sgn}(j-i)}{2^{i+j}i!j!}\,\left(s^2-1\right)^{\min(i,j)}\,\left[(s+1)^{\vert i-j\vert}-(s-1)^{\vert i-j\vert}\right]\times \\ &
\label{vapmoyalcomCG2} \times \left(\partial^{i}_{\bar z}\partial^{j}_{z}\,f\right)\,\left(\partial^{j}_{\bar z}\partial^{i}_{z}\,g\right)\, ,
\end{align}
\end{itemize}

\section{Interlude IV: quantization results with Born-Jordan weight}
\label{expcoefIV}
\begin{itemize}
  \item Born-Jordan weight: 
  
  \begin{equation}
\label{BJPi}
\Pi(q,p) = \dfrac{\sin qp}{qp}= \mathrm{sinc}(qp)\,, \quad \frac{1}{\Pi(q,p)} = qp \,\csc(qp)\,. 
\end{equation}
  \item Coefficients $\Dc$ are
  \begin{equation}
\label{CBJ}
\Dc_{kl}= \left\lbrace\begin{array}{cc}
    \delta_{kl}\, \dfrac{(-1)^r}{(2r+1)!}& \mbox{if}\ k=2r   \\
     0 &    \mbox{otherwise}
\end{array}\right.
\end{equation}
 \item Coefficients $\TDc$ are
 \begin{equation}
\label{TCBJ}
\TDc_{kl}= \left\lbrace\begin{array}{cc}
    \delta_{kl}\, \dfrac{(-1)^{r+1}\,2\left(2^{2r-1}-1\right)}{(2r)!}\,B_{2r}& \mbox{if}\ k=2r   \\
     0 &    \mbox{otherwise}
\end{array}\right.
\end{equation}
where $B_n$ is a Bell number, 
\begin{equation}
\label{BellN}
B_{n+1}= \sum_{k=0}^n \binom{n}{k}\, B_k\, , \quad B_0= 1\, .
\end{equation}
 \item Coefficients $\Dd$ are 
 \begin{equation}
\label{DBJ}
\Dd_{klk^{\prime}l^{\prime}}= \delta_{k+k^{\prime},l+l^{\prime}}\sum_{u,v}\binom{2u}{k^{\prime}-2v}\,\binom{2u}{k
l^{\prime}-2v}\, \Dc_{2u,2u}\, \TDc_{2v,2v}\, \TDc_{k+k^{\prime}-2(u+v),k+k^{\prime}-2(u+v)}\, ,
\end{equation}
if $k+k^{\prime}=l+l^{\prime}$ is even, otherwise is zero.
\item Etc. 


\end{itemize}

\section{Permanent issues of Weyl-Heisenberg integral quantizations with arbitrary weights}
\label{permissues}
\subsection{Quantization of linear and quadratic expressions in $z$ and $\bar z$ (or in $q$ and $p$)} 
\begin{itemize}
  \item CCR is a permanent outcome of the above quantization,
whatever the chosen complex function $\varpi\left(z\right)$, such that $\varpi\left(0\right)= 1$, provided
integrability and derivability at the origin is ensured.
  \item Quantization of canonical variables
  \begin{align}
\label{Azvp}
    A^{\vap}_{z}&=a-\left.\partial_{\bar{z}}\varpi\right\vert _{z=0}= a - c_{01}   \\
\label{Abzvp}  A^{\vap}_{\bar{z}} & =a^\dag+\left.\partial_{z}\varpi\right\vert _{z=0}= \adg + c_{10}\,,\\
\label{Aqvp} A^{\vap}_{q} & =\frac{1}{\sqrt{2}}\left[\left(a+a^\dag\right)-\left.\partial_{\bar{z}}\varpi\right\vert _{z=0}+\left.\partial_{z}\varpi\right\vert _{z=0}\right]= Q -\frac{1}{\sqrt{2}}\left[c_{01} - c_{10}\right]\,,\\
\label{Apvp} A^{\vap}_{p} & =\frac{1}{\sqrt{2}\ii}\left[\left(a-a^\dag\right)-\left.\partial_{\bar{z}}\varpi\right\vert _{z=0}-\left.\partial_{z}\varpi\right\vert _{z=0}\right] = P -\frac{1}{\sqrt{2}\ii}\left[ c_{01} + c_{10}\right]\,,
\end{align}
  \item Hence the  ccr, 
\begin{equation}
\label{ccrvp}
A^{\vap}_{q}A^{\vap}_{p}-A^{\vap}_{p}A^{\vap}_{q}=[Q,P]= i\left[a,a^\dag\right] = \ii\,I\,,
\end{equation}
\item Quadratic terms
\begin{align}
\label{Az2vp} A^{\vap}_{z^{2}} & =a^{2}-2a\left.\partial_{\bar{z}}\varpi\right\vert _{z=0}+\left.\partial_{\bar{z}}^{2}\varpi\right\vert _{z=0} = a^{2}-2c_{01}\,a  + 2 c_{02}\,,\\
\label{Abarz2vp} A^{\vap}_{\bar{z}^{2}} & =\left(a^\dag\right)^{2}+2a^\dag\left.\partial_{z}\varpi\right\vert _{z=0}+\left.\partial_{z}^{2}\varpi\right\vert _{z=0}= \left(\adg\right)^{2} +2c_{10}\,\adg  + 2 c_{20}\,,\\
\label{Azbarzvp} A^{\vap}_{\vert z \vert^{2}} & =a^\dag\,a +c_{10}\,a - c_{01}\,\adg  - c_{11} + \frac{1}{2}=a\,a^\dag +c_{10}\,a - c_{01}\,\adg  - c_{11} - \frac{1}{2}\,,
\end{align}
\begin{align}
\label{Aq2vp} A^{\vap}_{q^{2}} &= Q^2 + \sqrt{2}\,(c_{10}- c_{01})\,Q + c_{20} + c_{02} -c_{11}\, , \\
\label{Ap2vp} A^{\vap}_{p^{2}} &= P^2 + \sqrt{2}\,\ii\,((c_{10} + c_{01})\,P - c_{20} - c_{02} -c_{11}\, \\
\label{Aqpvp}  A^{\vap}_{qp} &= A^{\vap}_{q}A^{\vap}_{p}-\frac{\ii}{2} -\ii\,(c_{02} -c_{20}) + \frac{\ii}{2}\, \left(c_{01}^2 -  c_{10}^2\right)\\
\label{AqpvpQP} & = Q\,P + \frac{\ii}{\sqrt{2}}\,(c_{01} + c_{10})\,Q -\frac{1}{\sqrt{2}}\,(c_{01} - c_{10})\,P-\ii\,(c_{02} - c_{20})  - \frac{\ii}{2}\\
 & \equiv A^{\vap}_{q}A^{\vap}_{p}+(\mbox{constant}\in\mathbb{C})\,.
\end{align}
where the constant can take any value we wish depending on our choice of $\vap$ (think to the so-called $kp$-quantization!)
\item For more general formulae with variables $q$, $p$, and weight $\Pi(q,p)$, see Subsection \ref{monqp} below
\end{itemize}
\subsection{Quantization of arbitrary monomials in $z$ and $\bar z$}
\begin{itemize}
  \item Recurrence formula for $A^{\vap}_{z^n}$
  \begin{equation}
\label{recAzn}
A^{\vap}_{z^n}= \left(a-c_{01}\right)\,A^{\vap}_{z^{n-1}} - \sum_{\bar j =1}^{n-1}(-1)^{\bar j}a_{010\bar j}\,\frac{(n-1)!}{(n-1-\bar j)!}\, A^{\vap}_{z^{n-1-\bar j}}\, ,
\end{equation}
with for $\bar j \geq 1$
\begin{equation}
\label{a01bj}
a_{010\bar j} = d_{010\bar j}= \sum_{\bar k = 0}^{\bar j }(\bar j +1-\bar k) \, c_{0, \bar j  +1-\bar k}\,\tilde{c}_{0, \bar k}\, . 
\end{equation}
\item Recurrence formula for $A^{\vap}_{\bar z^n}$
  \begin{equation}
\label{recAbzn}
A^{\vap}_{\bar z^n}= \left(\adg + c_{10}\right)\,A^{\vap}_{\bar z^{n-1}} + \sum_{ j = 1}^{n-1}(-1)^j\,a_{10 j 0}\,\frac{(n-1)!}{(n-1- j)!}\, A^{\vap}_{\bar z^{n-1- j}}\, , 
\end{equation}
with for $ j \geq 1$
\begin{equation}
\label{a01bj}
a_{10 j 0} = d_{10 j 0} =  \sum_{ k = 0}^{j } (j+1-k)\, c_{ j+1 -k,0}\,\tilde{c}_{k ,0}\, . 
\end{equation}
  \item Separation formula for $A^{\vap}_{z^n\,\bar z^{\bar n}}$
  \begin{equation}
\label{sepAznbzbn}
A^{\vap}_{z^n\,\bar z^{\bar n}}=A^{\vap}_{ z^{n}} \,A^{\vap}_{\bar z^{\bar n}} + \sum_{ \bar i + j \geq1}(-1)^{\bar i}\,a_{0\bar i j 0}\,\frac{n!\,\bar n!}{(n-\bar i)\,(\bar n- j)!}\, A^{\vap}_{ z^{n-\bar i}} \,A^{\vap}_{\bar z^{\bar n -j}}\, .,
\end{equation}
with 
\begin{equation}
\label{ad0ij0}
a_{0\bar i j 0} = \sum_{ l=0}^{\min(\bar i,j)} \frac{1}{l!}\,\frac{1}{2^{l}}\, d_{0,\bar i - l, j-l,0 }\,  .
\end{equation}
\item It is easily inferred from these formula that 
\begin{itemize}
  \item $A^{\vap}_{z^n}$ is polynomial of degree $n$ in lowering operator $a$
  \begin{equation}
\label{Aznpola}
A^{\vap}_{z^n} = \sum_{m=0}^n \alpha_m^{\vap}\, a^m\, . 
\end{equation}
  \item $A^{\vap}_{\bar z^n}$ is polynomial of degree $n$ in raising operator $\adg$
    \begin{equation}
\label{Abznpolad}
A^{\vap}_{\bar z^n} = \sum_{m=0}^n \beta_m^{\vap}\, (\adg)^m\,.
\end{equation}
  \item $A^{\vap}_{z^n\,\bar z^{\bar n}}$ is polynomial in $a$ and $\adg$, separately of degree $n$ in $a$ and of degree $\bar n$ in $\adg$
\end{itemize} 
\end{itemize}

\subsection{Quantization of arbitrary monomials in $q$ and $p$ with weight $\Pi(q,p)$}
\label{monqp}
\begin{itemize}
\item Quantizations of $q$ and $p$
\begin{equation}
\label{Piqp}
A^{\Pi}_{q}= Q-\ii\,\Dc_{01}\, , \quad A^{\Pi}_{p}= P+\ii\,\Dc_{10}\, .
\end{equation}
 \item Recurrence formula for $A^{\Pi}_{q^n}$
  \begin{equation}
\label{recAqn}
A^{\Pi}_{q^n}= \left(Q-\ii\,\Dc_{01}\right)\,A^{\Pi}_{q^{n-1}} + \sum_{l^{\prime}=1}^{n-1}(-\ii)^{l^{\prime}+1}\Da_{010l^{\prime}}\,\frac{(n-1)!}{(n-1-l^{\prime})!}\, A^{\Pi}_{q^{n-1-l^{\prime}}}\, ,
\end{equation}
with for $l^{\prime} \geq 1$
\begin{equation}
\label{a01bjqp}
\Da_{010l^{\prime}} = \Dd_{010l^{\prime}}= \sum_{s= 1}^{l^{\prime} +1 }s \, \Dc_{0 s}\,\TDc_{0, l^{\prime} +1-s}\, . 
\end{equation}
\item Recurrence formula for $A^{\Pi}_{p^n}$
  \begin{equation}
\label{recApn}
A^{\Pi}_{p^n}= \left(P + \ii\,\Dc_{10}\right)\,A^{\Pi}_{p^{n-1}} + \sum_{ k^{\prime} = 1}^{n-1}(\ii)^{k^{\prime}+1}\,\Da_{10 k^{\prime} 0}\,\frac{(n-1)!}{(n-1- k^{\prime})!}\, A^{\Pi}_{p^{n-1- k^{\prime}}}\, , 
\end{equation}
with for $ k^{\prime}\geq 1$
\begin{equation}
\label{a01bjqp}
\Da_{10 k^{\prime} 0} = \Dd_{10 k^{\prime} 0} =  \sum_{ r = 1}^{k^{\prime}+1} r\, \Dc_{r0}\,\TDc_{k^{\prime}+1-r ,0}\, . 
\end{equation}
  \item Separation formula for $A^{\Pi}_{q^m\,p^n}$
  \begin{equation}
\label{sepAqnpn}
A^{\Pi}_{q^m\,p^n}=A^{\Pi}_{ q^{m}} \,A^{\Pi}_{p^{n}} + \sum_{ l+ k^{\prime} \geq1}\ii^{k^{\prime}-l}\,a_{0k^{\prime} l 0}\,\frac{m!\,n!}{(m-l)\,(n- k^{\prime})!}\, A^{\Pi}_{ q^{m-l}} \,A^{\Pi}_{p^{n -k^{\prime}}}\, .,
\end{equation}
with 
\begin{equation}
\label{ad0ij0qp}
\Da_{0 l k^{\prime} 0} = \sum_{ s=0}^{\min(l,k^{\prime})} \frac{(-\ii)^s}{s!}\,\frac{1}{2^{s}}\, \Dd_{0,l- s, k^{\prime}-s,0 }\,  .
\end{equation}
\item It is easily inferred from these formula that 
\begin{itemize}
  \item $A^{\Pi}_{q^n}$ is polynomial of degree $n$ in position operator $Q$
  \begin{equation}
\label{Aqnpola}
A^{\Pi}_{q^n} = \sum_{m=0}^n u_m^{\Pi}\, Q^m\, . 
\end{equation}
  \item $A^{\Pi}_{p^n}$ is polynomial of degree $n$ in momentum operator $P$
    \begin{equation}
\label{Apnpolad}
A^{\Pi}_{p^n} = \sum_{m=0}^n v_m^{\Pi}\, P^m\,.
\end{equation}
  \item $A^{\Pi}_{q^m\,p^{n}}$ is polynomial in $Q$ and $P$, separately of degree $m$ in $Q$ and of degree $n$ in $P$
\end{itemize} 
\end{itemize}

\section{Quantizations with particular weight functions \\
(superscript $\vap$ is omitted)}
\label{partweights}
\begin{itemize}
 \item  Regular quantizations\\
 The weight function $\varpi$ is even and real, $\varpi(-z)=\varpi(z)$,  $\overline{\varpi(z)}=\varpi(z)$ 
\begin{equation}
\label{regquant1}
A_{z} = a\, , \quad  A_{\overline{f(z)}} = A_{f(z)}^\dag\, . 
\end{equation}

\item Elliptic regular quantizations\\
The weight function $\varpi$ is isotropic $\varpi(z) \equiv w(|z|^2)$ with $w: \R \mapsto \R$ \\
 Example: Cahill-Glauber choice
\begin{equation}
\label{standvarpi}
\varpi_s(z) = e^{s |z|^2/2}\, , \quad \mathrm{Re}\; s<1\, . 
\end{equation}
From
\begin{equation}
\label{integasslag}
\int_{0}^{\infty} \, e^{-\nu x}\, x^{\lambda}\, L_{n}^{\alpha}(x)\ud x =\frac{\Gamma(\lambda + 1)\Gamma(\alpha+n+1)}{n!\, \Gamma(\alpha + 1)}\nu^{-\lambda -1 }{}_{2}F_1(-n,\lambda + 1; \alpha + 1; \nu^{-1})\, , 
\end{equation}
we get the diagonal elements of  ${\sf M}_s$:
\begin{equation}
\label{diagMs}
\lg e_n|{\sf M}_s|e_n\rg = \frac{2}{1-s}\,\left( \frac{s+1}{s-1} \right)^n\, , 
\end{equation}
and so
\begin{equation}
\label{defMs}
{\sf M}_s= \int_{\mathbb{C}}\;\varpi_s(z) D(z) \,\frac{{\ud}^2z}{\pi }= \frac{2}{1-s} \exp \left\lbrack\left( \log \dfrac{s+1}{s-1}\right) a^\dag a \right\rbrack\,.
\end{equation}
 $s=-1$ corresponds to the CS  (anti-normal) quantization, since 
\begin{equation*}
{\sf M}_{-1}= \lim_{s\to -1_{-}} \dfrac{2}{1-s} \exp \left\lbrack\left( \ln \dfrac{s+1}{s-1}\right) a^\dag a \right\rbrack = |e_0\rg\lg e_0|\, , 
\end{equation*}
and so 
\begin{equation}
\label{csquants-1}
A_f = \int_{\mathbb{C}} \,  D(z){\sf M}_{-1}D(z)^{\dag} \,f(z) \, \frac{\ud^2 z}{\pi}= \int_{\mathbb{C}} \,  |z\rg\lg z| \, f(z) \,\,\frac{\ud^2 z}{\pi}\, .
\end{equation}
$s=0$ corresponds to the Weyl-Wigner quantization since, from Eq. (\ref{defMs}), 
$
\sf M_0= 2\sf P
$,
and so 
\begin{equation}
\label{wigweylquant}
A_f = \int_{\mathbb{C}} \, D(z) \,2{\sf P}\, D(z)^{\dag}\,  f(z) \,  \,\frac{\ud^2 z}{\pi}\, .
\end{equation}
$s=1$ is the normal quantization in an asymptotic sense. \\
The operator ${\sf M}_s$ is positive unit trace class for $s \leq -1$ (it is just trace class if $\mathrm{Re}\;s<0$)

\item Hyperbolic regular quantizations\\
The weight function $\varpi$ verifies $\varpi(z)\equiv
 {\sf m}(\mathrm{Im}\; (z^2))$ with ${\sf m}: \R \mapsto \R$ (i.e. yields a regular quantization), like the Born-Jordan weight. \\
It is easily infererred from \eqref{Iquq} and \eqref{Iqvp} that  this case preserves the classical functions 
\begin{equation}
A_{f(q)}=f(Q)\, , \quad A_{f(p)} = f(P) \,.
\end{equation}
Because $\mathrm{Im}\; (z^2) \equiv q\, p$, we only need the Planck constant $\hbar$ and a mathematical function ${\sf m}$ to build the physical expression  ${\sf m}(\mathrm{Im}\; (z^2)/\hbar)$ \\
\item The common elliptic regular and hyperbolic regular cases must verify $\varpi(z) = w(|z|^2) =   {\sf m}(\mathrm{Im}\; (z^2))$ with $\varpi(0)=1$. The unique solution is $\varpi(z)=1$ corresponds to the Weyl-Wigner quantization

\item Isometric quantizations\\
If  we have
\begin{equation}
\mathrm{tr} (A_f^\dag A_f) = \int_{\mathbb{C}} |f(z)|^2 \frac{\ud^2z}{\pi}\, .
\end{equation}
 $f \mapsto A_f$ is then invertible (the inverse is given by a trace formula)\\
From \eqref{traceDD1} we have the trace formula
\begin{equation}
\mathrm{tr} (A_f^\dag A_f) = \int_{\C} \frac{\ud^2z}{\pi} |\varpi(z)|^2 \vert\omfs(z)\vert^2 \,.
\end{equation}
From the invariance of the $L^2$-norm under symplectic transform, we find that $f \mapsto A_f$ is isometric if and only if $| \varpi(z) |=1$ for all $z$

\item Elliptic regular quantizations that are isometric\\
It is the case $\varpi(z)=w(|z|^2) \in \{ -1,+1 \}$. 
Simple example
\begin{equation}
\varpi_\alpha(z)=2 \theta(1-\alpha |z|^2)-1
\end{equation}
where $\theta$ is the Heaviside function. The Weyl-Wigner quantization is a special case ($\alpha=0$) 

\item Hyperbolic regular quantizations that are isometric\\
It is the case ${\sf m}(u) \in \{ -1,+1 \}$. \\
Simple example
\begin{equation}
\varpi_\alpha(z)=2 \theta(1-\alpha \mathrm{Im}\; (z^2))-1\,. 
\end{equation}
The Weyl-Wigner quantization is a special case ($\alpha=0$) 
\end{itemize}

\section{Separable quantizations}
\label{sepquant}
\begin{itemize}
  \item It is the case $\Pi(q,p)= \lambda(q)\, \mu(p)$ \underline{and} $F(q,p) = L(q)\,M(p)$. From \eqref{quantPi2}
\begin{equation}
\label{sepquant1}
A_{L(q)\,M(p)}= \int_{\R^2}\frac{\ud q\,\ud p}{2\pi}\, e^{-\ii \frac{qp}{2}}\, e^{\ii pQ}\, e^{-\ii qP}\, \lambda(q) \, \mu(p)\, \mfF[L](p)\,\mfF[M](-q)\,. 
\end{equation}
  \item Useful case $M(p)= p^m$ whereas $L(q)$ is a function  of class $C^m$ (actually all this may be extended to distributions with appropriate weight functions)
  \begin{equation}
\label{sepquant2}
A_{L(q)\,p^m}=  \sum_{\substack{
         r,s,t\\
         r+s+t=m}}  2^{-s}\, \binom{m}{r\,s\,t}\,\ii^r\,\lambda^{(r)}(0)\,(-i)^s \frac{1}{\sqrt{2\pi}}\,\left(\overline{\mfF}[\mu]\ast L^{(s)} \right)(Q)\, P^t\,,
\end{equation}
where ``$\ast$'' is the convolution product
\begin{equation}
\label{convdef}
f\ast g(x) = \int_{-\infty}^{+\infty}f(x-y)\,g(y)\,\ud y = g\ast f(x) \, , 
\end{equation}
and $\overline{\mfF}$ is the inverse Fourier transform (see \eqref{fourdist1} and  \eqref{fourdist2}) 
\item Weyl-Wigner case ($\lambda(q)= 1=\mu(p)$) for $M(p)= p^m$ and $L(q)$ is a function  of class $C^m$
  \begin{equation}
\label{sepquantWW}
A^{\mathcal{W-W}}_{L(q)\,p^m}= \sum_{t} 2^{t-m}\, \binom{m}{t}\, (-\ii)^{m-t}\,  L^{(m-t)}(Q)\, P^t\,.
\end{equation}
and this corresponds to a certain type of quantization (``Weyl calculus'')
\item Lowest $m$'s
\begin{equation}
\label{m0}
  A^{\mathcal{W-W}}_{L(q)}    = L(Q)\, , 
\end{equation}
\begin{equation}
\label{m1}
A^{\mathcal{W-W}}_{L(q)\,p}     = \frac{1}{2}\, (-\ii)\,L^{\prime}(Q) + L(Q)\, P = \frac{1}{2}[ L(Q) \,P + P\,L(Q)]\, , 
\end{equation}

\begin{align}
\nonumber A^{\mathcal{W-W}}_{L(q)\,p^2} &=- \frac{1}{4}\,L^{\prime\prime} (Q)- \ii\, L^{\prime}(Q)\, P + L(Q)\, P^2\\
\label{m2} & =  \frac{1}{12}\, L^{\prime\prime}(Q)  + \frac{1}{3}\left(P^2\,L(Q) + P\,L(Q)\,P + L(Q)\,P^2\right) \equiv \frac{1}{12}\, L^{\prime\prime}(Q) + \mathrm{Sym}_{\mathrm{W}}\left(L(Q)\,P^2\right)
\end{align}
\begin{align}
\label{m3} A^{\mathcal{W-W}}_{L(q)\,p^3} &= \frac{-\ii }{8}\, L^{\prime\prime\prime} + \frac{7}{4}\, L^{\prime\prime}\,P
+ \mathrm{Sym}_{\mathrm{W}}\left(L\,P^3\right)\, , \\
\nonumber & \mathrm{etc}\, , 
 \end{align}
where $ \mathrm{Sym}_{\mathrm{W}}$ stands for the so-called Weyl ordering. The latter is defined for monomials with powers of two operators $A$ and $B$ as
\begin{equation}
\label{weylcalculus}
 \mathrm{Sym}_{\mathrm{W}}\left(A^mB^n\right)=\frac{1}{\binom{m+n}{m}} \sum_{C_1 \cdots  C_{m+n}}C_1 \cdots C_{m+n}
\end{equation} 
where the $C_1\cdots C_{m+n}$ ranges over all uples which contain $m$ copies of $A$ and $n$ copies of $B$
\item Ingredients of the proof of \eqref{sepquant1}-\eqref{sepquantWW} from basic Fourier analysis and distribution theory 
\begin{align}
\label{fourdist1}
   \mfF[f](k) &:= \frac{1}{\sqrt{2\pi}} \int_{-\infty}^{\infty} f(x)\,e^{-\ii kx}\, \ud x\, , \quad \overline{\mfF}[f](k) :=  \mfF[f](-k) =\frac{1}{\sqrt{2\pi}} \int_{-\infty}^{\infty} f(x)\,e^{\ii kx}\, \ud x\, ,\\
 \label{fourdist2}  \overline{\mfF}\,\mfF &= \mfF\,\overline{\mfF}= I\\
 \label{fourdist3}  \lg \delta^{(m)}\, , \, \varphi\rg &\equiv \int_{-\infty}^{+\infty}(-1)^m \delta^{(m)}(x)\,\varphi(x)\, \ud x= \varphi^{(m)}(0)\, , \\
  \label{fourdist4} \mfF\left[x^m\right]&= \sqrt{2\pi}\, \ii^m\,\delta^{(m)}\\
  \label{fourdist5}  \mfF[f \ast g](k) = & \sqrt{2\pi}\mfF[f](k)\,(\mfF[g](k) \, , \quad 
 \mfF[F(k)\, G(k) ] =   \frac{1}{\sqrt{2 \pi}}\mfF[F] \ast \mfF[G]\,.
\end{align}
\end{itemize}

\section{Useful commutation relations}
\label{comrules}
\begin{itemize}
  \item Restoring the presence of $\hbar$ and starting from $[Q,P]= \ii \hbar\,I$,
  \item Commutator of $P$ with powers of $Q$
  \begin{equation}
\label{pqm}
[P,Q^m]  = -m \ii \hbar Q^{m-1} 
\end{equation}
\item Commutator of $P^2$ with powers of $Q$
\begin{equation}
\label{p2qm}
[P^2,Q^m]  =  (-\ii \hbar)^2 m (m-1) Q^{m-2} - 2 m \ii \hbar Q^{m-1} P 
\end{equation}
\item Commutator of powers of $P$ with $Q$
\begin{equation}
\label{p2qm}
[P^n,Q]  =  - n i \hbar P^{n-1}  
\end{equation}
\item Commutator of powers of $P$ with $Q^2$
\begin{equation}
\label{p2qm}
[P^n,Q^2]  = (-i \hbar)^2 n (n-1) P^{n-2}  -2 n i \hbar Q P^{n-1}  
\end{equation}

\item Commutator of $P^m$ with $U(Q)$, $U$ of class $C^m$ (at least)
\begin{equation}
\label{puq}
[P,U(Q)] = -\ii \hbar U^\prime (Q) \, ,
\end{equation}

\begin{equation}
\label{pmuq}
[P^m,U(Q)] = \sum_{k=0}^{m-1} \binom{m}{k} (-i \hbar)^{m-k} \,U^{(m-k)} (Q) \, P^k \, ,
\end{equation} 
%
\item If we consider powers of $P$ and $Q$ we have: 
\begin{enumerate}
  \item[(i)] For $n \leq m$
\begin{equation}
\label{pnqm}
[P^n, Q^m] = \sum_{k=0}^{n-1} (- i \hbar)^{n-k} \binom{n}{k} 
A_{m, n-k} Q^{m-n+k} P^k \, ,
\end{equation}
where $A_{m, n} = m! / n!$ is the arrangement of $m$ and $n$
  \item[(ii)] For $n \geq m$
\begin{equation}
\label{pnqm}
[P^n, Q^m] = \sum_{k=0}^{m-1} (- i \hbar)^{m-k} \binom{m}{k} 
A_{n, m-k} Q^{k} P^{n-m+k} \, ,
\end{equation}
\end{enumerate}
\end{itemize}

\section{Gaussian separable quantizations}
\label{gausssepquant}
\begin{itemize}
  \item It is the case when the weight is 
 \begin{equation}
\label{gaulamu}
 \vp(z) = \Pi(q,p) = \lambda(p)\, \mu(p) = e^{-\frac{q^2}{2\sigma_{\ell}^2}}\, e^{-\frac{p^2}{2\sigma_{\eth}^2}}\,.
\end{equation} 
  \item This case includes all Cahill-Glauber weights $\vp(z)= e^{s\frac{z^2}{2}}$, and in particular the ``limit'' Weyl-Wigner 
  as both the widths $\sigma_{\ell}$ and $\sigma_{\eth}$ are infinite (Weyl-Wigner is singular in this respect!)
  \item Specifying \eqref{sepquant2}  and taking into account
  \begin{equation}
\label{lambdas}
\begin{split}
\lambda^{(r)}(0)& = (-1)^r\,(2)^{-r/2}\,(\sigma_{\ell})^{-r} \,H_r(0)\underset{r=2u}{=}\frac{1}{\sqrt{\pi}}\,\left(\frac{-2}{\sigma^2_{\ell}}\right)^u\, \Gamma\left(u+ \frac{1}{2}\right)\\
& \underset{r=2u+1}{=} 0
\end{split}
\end{equation}
and 
\begin{equation}
\label{fourmu}
\overline{\mfF}[\mu](x)= \sigma_{\eth}\, e^{-\sigma_{\eth}^2\frac{x^2}{2}}\, ,
\end{equation}
we have 
\begin{equation}
\label{sepquantgauss}
\begin{split}
A^{\mathcal{Gauss}}_{L(q)\,p^m}&= \sum_{\substack{
         u,s,t\\
         2u+s+t=m}}  2^{u-s}\, \binom{m}{2u\,s\,t}\,\sigma^{-2u}_{\ell}\,  \frac{\Gamma(u+ 1/2)}{\sqrt{\pi}}\,(-\ii)^s \,\mathfrak{G}_{1/\sigma_{\eth}}\left[L^{(s)}\right](Q)\, P^t\\
 &=    \sum_{\substack{
         u,s,t\\
         2u+s+t=m}}  2^{u-s}\, \binom{m}{2u\,s\,t}\,\sigma^{-2u}_{\ell}\,  \frac{\Gamma(u+ 1/2)}{\sqrt{\pi}}\,(-\ii)^s \,\frac{d^s}{dQ^s}\mathfrak{G}_{1/\sigma_{\eth}}\left[L\right](Q)\, P^t\, ,      
\end{split}
\end{equation}
where $\mathfrak{G}_{\sigma}[F]$ stands for the Gaussian convolution of $F$,
\begin{equation}
\label{gaussconv}
\mathfrak{G}_{\sigma}[F](x):= \frac{1}{\sigma \sqrt{2\pi}}\int_{-\infty}^{+\infty}\ud y\, e^{-\frac{(x-y)^2}{2\sigma^2}}\, F(y)\,. 
\end{equation}
\end{itemize}

\subsection*{Particular cases of Gaussian separable quantizations}
\begin{itemize}
\item Kinetic energy with ``variable mass'' : ${\sf m} = {\sf m}(q) = 1/(2L(q))$
\begin{align}
\label{Lqp2}
A^{\mathcal{Gauss}}_{L(q)\,p^2}&= \mathfrak{G}_{1/\sigma_{\eth}}\left[L\right](Q)\, P^2 - \ii\,\mathfrak{G}_{1/\sigma_{\eth}}\left[L^{\prime}\right](Q)\, P + \mathfrak{G}_{1/\sigma_{\eth}}\left[\frac{L}{\sigma^2_{\eth}} - \frac{L^{\prime\prime}}{4}\right](Q)\\
&= P\, \mathfrak{G}_{1/\sigma_{\eth}}\left[L\right](Q)\, P  + \mathfrak{G}_{1/\sigma_{\eth}}\left[\frac{L}{\sigma^2_{\eth}} - \frac{L^{\prime\prime}}{4}\right](Q) \, . 
\end{align}
  \item $L(q)= 1$, $M(p)=p^m$
 \begin{equation}
\label{L1pm}
  A^{\mathcal{Gauss}}_{p^m}  =  \sum_{u=0}^{\left\lfloor \frac{m}{2}\right\rfloor} \frac{1}{2^u\,\sigma_{\ell}^{2u}}\, \frac{m!}{u!\,(m-2u)!}\, P^{m-2u}= 2^{-m/2}\,\sigma_{\ell}^{-m}\,(-\ii)^m\,H_m\left(\frac{i\sigma_{\ell}\,P}{\sqrt{2}}\right)\, , 
\end{equation}   
 \begin{align}
  \label{L2p1}   A^{\mathcal{Gauss}}_{p}&=  P\\
  \label{L3p2} A^{\mathcal{Gauss}}_{p^2}& = P^2 + \frac{1}{\sigma_{\ell}^2}\,. 
   \end{align}
   where $H_m(x)$ is a Hermite polynomial
 \item $L(q)$, $M(p)=1$
 \begin{equation}
\label{LqM1}
   A^{\mathcal{Gauss}}_{L(q)}  = \mathfrak{G}_{1/\sigma_{\eth}}\left[L\right](Q)= \frac{\sigma_{\eth}}{\sqrt{2\pi}}\int_{-\infty}^{+\infty} \ud y \, e^{-\sigma_{\eth}^2\frac{y^2}{2}}\, L(Q-y)\,,\\
\end{equation}
 \begin{equation}
\label{LqmM1} A^{\mathcal{Gauss}}_{q^m} = \sum_{u=0}^{\left\lfloor \frac{m}{2}\right\rfloor} \frac{1}{2^u\,\sigma_{\eth}^{2u}}\, \frac{m!}{u!\,(m-2u)!}\, Q^{m-2u}=2^{-m/2}\,\sigma_{\eth}^{-m}\,(-\ii)^m\,H_m\left(\frac{i\sigma_{\eth}\,Q}{\sqrt{2}}\right)\, ,
\end{equation}
  \begin{align}  
   \label{Lq1M1}   A^{\mathcal{Gauss}}_{q}&=  Q\,,\\
  \label{Lq2M1} A^{\mathcal{Gauss}}_{q^2}& = Q^2 + \frac{1}{\sigma_{\eth}^2}\,.
      \end{align}
\item Harmonic oscillator
\begin{equation}
\label{Lq2+Mp2}
 A^{\mathcal{Gauss}}_{\frac{p^2+q^2}{2}} = \frac{P^2+Q^2}{2} + \frac{1}{2}\left(\frac{1}{\sigma_{\ell}^2} + \frac{1}{\sigma_{\eth}^2}\right)= N + \frac{1}{2}\left(\frac{1}{\sigma_{\ell}^2} + \frac{1}{\sigma_{\eth}^2} + 1\right)\,.
\end{equation}
\item Comparing with the CS quantization (Cahill-Glauber $s=-1$) of the harmonic oscillator \eqref{quantosc2} below ($\sigma_{\ell} = \sigma_{\eth} =\sqrt{2}$), we infer that the CS case (``zero temperature'') and separable Gaussian case yield same HO eigenenergies when 
\begin{equation}
\label{csgauss}
\frac{1}{\sigma_{\ell}^2} + \frac{1}{\sigma_{\eth}^2}= 1\,, 
\end{equation}
\item and more generally the ``$s$'' Cahill-Glauber quantization ($\sigma_{\ell} = \sigma_{\eth} = \sqrt{-2/s}$) and  separable Gaussian quantizations yield same HO eigenenergies when 
\begin{equation}
\label{CHgauss}
\frac{1}{\sigma_{\ell}^2} + \frac{1}{\sigma_{\eth}^2}= -s\,. 
\end{equation}
\end{itemize}

\section{Quantum harmonic oscillator according to regular $\varpi$}
\label{QOintq}
Given a general weight function $\varpi$, we saw in the above that the quantization of the classical harmonic oscillator energy $\vert z \vert^2 = (p^2 + q^2)/2$ yields the operator
\begin{equation}
\label{quantosc1}
A^{\vap}_{\vert z \vert^2} =  \adg a + \left.\partial_{z}\varpi\right\vert_{z=0}\,a -  \left.\partial_{\bar z}\varpi\right\vert_{z=0}\,\adg + \frac{1}{2} -  \left.\partial_{z}\partial_{\bar z}\varpi\right\vert_{z=0}\,.
\end{equation}
In the case of a regular quantization, i.e., $\varpi(-z)=\varpi(z)$,  $\overline{\varpi(z)}=\varpi(z)$, we obtain the operator
\begin{equation}
A^{\vap}_{\vert z \vert^2} = \adg a + \frac{1}{2}  -  \left.\partial_{z}\partial_{\bar z}\varpi\right\vert_{z=0}\,,
\end{equation}
and for $q^2$ and $p^2$,
\begin{eqnarray}
A^{\vap}_{q^2} = Q^2 - \left.\partial_{z}\partial_{\bar z}\varpi\right\vert_{z=0} + \frac{1}{2} \left( \left.\partial_{z}^2 \varpi\right\vert_{z=0}+ \left.\partial_{\bar z}^2 \varpi\right\vert_{z=0}\right) \\
A^{\vap}_{p^2}= P^2 - \left.\partial_{z}\partial_{\bar z}\varpi\right\vert_{z=0} - \frac{1}{2} \left( \left.\partial_{z}^2 \varpi\right\vert_{z=0}+ \left.\partial_{\bar z}^2 \varpi\right\vert_{z=0}\right)
\end{eqnarray}

One observes that the difference between the ground state energy of the quantum harmonic oscillator, namely $E_0= 1/2- \left.\partial_{z}\partial_{\bar z}\varpi\right\vert_{z=0}$, and the minimum of the \underline{quantum} potential energy, namely 
$$E_m=\frac{1}{2}\,[\min(A^{\vap}_{q^2}) + \min(A^{\vap}_{p^2})] = - \left.\partial_{z}\partial_{\bar z}\varpi\right\vert_{z=0}$$
  is $E_0-E_m=1/2$. It is independent of the particular (regular) quantization chosen 

In the exponential Cahill-Glauber case $\varpi_s(z) = e^{s\vert z \vert^2/2}$ the above operators reduce to 
\begin{equation}
\label{quantosc2}
 A^{\vap_s}_{q^2}=Q^2-\frac{s}{2} \,, \quad A^{\vap_s}_{p^2}=P^2 - \frac{s}{2}\,, \quad A^{\vap_s}_{\vert z \vert^2} = \adg a + \frac{1-s}{2} \,.
\end{equation}

\section{Variations on the Wigner function and lower symbols}
\label{wignerandothers}
\begin{itemize}
 \item The Wigner function is (up to a constant factor) the Weyl transform of a quantum density operator. For a particle in one dimension it takes the form (in units $\hbar=1$)
\begin{equation}
\label{wigtransf1}
\mathsf{W}(q,p)= \frac{1}{2\pi}\int_{-\infty}^{+\infty}\, \left\lg q-\frac{y}{2}\right\vert \rho \left\vert q +\frac{y}{2}\right\rg\, e^{\ii py}\, \ud y\,. 
\end{equation}
\item Adapting this definition to the present context, and  given an operator $A$, the corresponding  Wigner function is defined as the linear map
\begin{equation}
\label{wigA}
A \mapsto\mathfrak{W}^{\mathcal{W-W}}_A(z) = \mathrm{tr}\left(D(z)2{\sf P}D(z)^{\dag}A\right)\, ,
\end{equation}
\item In the case of the quantization map $f \mapsto A_f$ based on a weight function $\varpi$, we get the ``lower symbol''  $\check{f}$ of $A_f$
\begin{align}
\label{lowsymb1}
\mathfrak{W}^{\vap}_{A_f}(z):= \mathrm{tr}\left(\sfM^{\vap}(z)A_f\right)  \equiv \check{f}(z)&=  \int_{\C} \,  \mfs\left[\varpi\,\widetilde\varpi\right](\xi-z)\, f(\xi)\, \frac{\ud^2 \xi}{\pi}\\
\label{lowsymb2} &=  \int_{\C} \,  \mfs\left[\varpi\right]\,\ast \,\mfs\left[\widetilde\varpi\right](\xi-z)\, f(\xi)\, \frac{\ud^2 \xi}{\pi^2}\\
\label{lowsymb3} &=  \mfs\left[\frac{\varpi}{\pi}\right]\ast \omfs\left[\frac{\varpi}{\pi}\right]\ast f(z)\\
\label{lowsymb4} &= \int_{\C} \,\varpi(\xi)\,\varpi(-\xi)\, \omfs\left[\mathsf{t}_{-z}\,f\right](\xi)\, \, \frac{\ud^2 \xi}{\pi}\,. 
\end{align}
\item Hence the map $f\mapsto  \check{f}$ is an (Berezin-like) integral transform with kernel $\mfs\left[\varpi\,\widetilde\varpi\right](\xi-z)$
\item From \eqref{lowsymb3} it is also viewed as the convolution of $f$ with the correlation of the (quasi-) distributions $z\mapsto  \mfs\left[\frac{\varpi}{\pi}\right](z)$ and $z\mapsto  \omfs\left[\frac{\varpi}{\pi}\right](z)=\mfs\left[\frac{\varpi}{\pi}\right](-z)$  on the phase space $\C$ 
\item If the latter are non-negative, i.e. are both probability distribution on $\C$ equipped with the uniform measure $\ud^2 z$,  their convolution product is the probability distribution for the random variable $z_1-z_2$
\item Then the map $f\mapsto  \check{f}$ is interpreted as an averaging 
\item Hence we reach here the most sensitive aspect of Weyl-Heisenberg integral quantization through the combination  of integrals, symmetry and probabilities, see \cite{gazeau_intsympro16}
\item In the case of Weyl-Wigner quantization
\begin{equation}
\label{wigAW}
\mathfrak{W}^{\mathcal{W-W}}_{A_f} =  \check{f} =f
\end{equation}
(this one-to-one correspondence of the Weyl quantization is related to the isometry property) 

\item In the case of the anti-normal quantization, the above convolution corresponds to the Husimi transform (when $f$ is the Wigner transform of a quantum pure state)

\item If the quantization map $f \mapsto A_f$ is regular (i.e. $\varpi$ real even) and isometric (i.e. $\vert \varpi\vert = 1$), which means  that $\varpi(z)$ is even and $\in \{-1,1\}$,   the corresponding inverse map $A \mapsto \mathfrak{W}_A$ is given by
\begin{equation}
\label{invmapquant}
\mathfrak{W}^{\vap}_A (z)= \mathrm{tr} \left(D(z) {\sf M} D(z)^{\dag}A\right)\, , \ \mathrm{where} \ {\sf M} =  {\sf M}^\dag= \int_{\mathbb{C}} \, \varpi(z) D(z)\, \dfrac{\ud^2 z}{\pi} \, . 
\end{equation}
\item In general this map $A \mapsto \mathfrak{W}^{\vap}_A$ is only the dual of the quantization map $f \mapsto A_f$ in the sense that
\begin{equation}
\int_{\mathbb{C}} f(z)\,\mathfrak{W}^{\vap}_A(z) \,\frac{\ud^2z}{\pi}= \mathrm{tr}( A\, A_f)\,.
\end{equation}
\item This dual map becomes the inverse of the quantization map only in the case of a Hilbertian isometry
\end{itemize}

\section{Conclusion}

It is clear that no compendium is perfect and complete. The hope of this one is to live its own life, being collectively improved with corrections, additions, suggestions. We invite interested researchers to be part of the work  in proposing  apposite remarks or contributions. 

\end{document}